\documentclass[12pt,onecolumn,journal]{IEEEtran}
\usepackage{url,amsmath,graphicx}
\usepackage{amsfonts}
\usepackage{dblfloatfix}
\usepackage{multirow}
\usepackage{amssymb}
\usepackage{mathtools}
\usepackage{amsmath}
\usepackage{relsize}
\usepackage{geometry}
\usepackage[T1]{fontenc}
  
\usepackage{bm}
\usepackage{pifont}
\usepackage{color}
\DeclareMathOperator{\argmax}{argmax}
\begin{document}
\title{Time Synchronization of Turbo-Coded \vspace {-0.5cm}Square- \vspace {-0.5cm} QAM-Modulated Transmissions: Code-Aided ML Estimator and Closed-Form Cram\'er-Rao Lower\vspace {-0.5cm} Bounds} 
\author{ Faouzi Bellili$^*$, Achref Methenni, Souheib Ben Amor, Sofi\`ene Affes, and Alex Stéphenne
\\\small INRS-EMT, 800, de la Gaucheti\`ere Ouest, Bureau 6900, Montreal, Qc, H5A 1K6, Canada.
   \\\small Emails: \{bellili, methenni, souheib.ben.amor, affes\}@emt.inrs.ca, and stephenne@ieee.org 
\vspace{0.3cm}
\thanks{This work was submitted to IEEE Transactions on signal Processing. Copyright (c) 2015 IEEE. Personal use of this material is permitted. However, permission to use this material for any other purposes must be obtained from the IEEE by sending a request to pubspermissions@ieee.org. This work was supported by a Canada Research Chair in Wireless Communications and a Discovery Accelerator Supplement from NSERC. Work accepted for publication, in part, in IEEE Globecom 2015.
}
}
\maketitle
\pagenumbering{arabic}
 \vspace {-3cm}
\begin{abstract}  
This paper introduces  a new  maximum likelihood (ML) solution for the
code-aided (CA) timing recovery problem in square-QAM transmissions and derives, for the very first time, its CA Cramér-Rao lower bounds (CRLBs) in closed-form expressions.  By exploiting the full symmetry  of square-QAM constellations and further scrutinizing the Gray-coding mechanism, we  express  the likelihood function (LF) of the system explicitly in terms of the code bits' \textit{a priori} log-likelihood ratios (LLRs). The timing recovery task is then embedded in the turbo iteration loop wherein increasingly accurate estimates for such LLRs are computed from the output of the soft-input soft-output (SISO) decoders and exploited at a per-turbo-iteration basis in order to refine the ML time delay  estimate. The latter is then  used to better re-synchronize the system, through feedback to the matched filter (MF),  so as to obtain more reliable symbol-rate samples for the next turbo iteration. In order to properly benchmark the new CA ML estimator, we also derive for the very first time the closed-form expressions for the exact CRLBs of the underlying turbo synchronization problem. Computer simulations  will show  that the new closed-form CRLBs coincide exactly with their empirical counterparts evaluated previously using exhaustive Monte-Carlo simulations. They will also show unambiguously the remarkable performance improvements of CA estimation against the traditional non-data-aided (NDA) scheme; thereby  highlighting  the potential performance gains in time synchronization that can be achieved owing to the decoder assistance.  Over a wide range of practical  SNRs, CA estimation  becomes even equivalent to the \textit{completely} data-aided (DA)  scheme in which all the transmitted symbols are perfectly known to the receiver. Moreover, the new CA ML estimator almost reaches the underlying CA CRLBs, even for small SNRs, thereby confirming its statistical efficiency in practice. It also enjoys significant improvements in computational complexity as compared to the most powerful existing ML solution, namely the combined sum-product and expectation-maximization (SP-EM) algorithm.
\end{abstract}

\section{Introduction}
\IEEEPARstart{I}{n} order to provide high quality of service while satisfying the ever-increasing demand in high data
 rates,  the use of powerful error-correcting codes in conjunction with  high-spectral-efficiency modulations  is 
 advocated. Indeed, turbo codes along with  high-order quadrature amplitude modulations (QAMs) are two key features
 of current and future wireless communication standards such as 4G long-term evolution (LTE), LTE-advanced (LTE-A)
 and beyond (LTE-B) [\ref{LTE1}, \ref{LTE2}]. As a crucial task in any digital receiver [\ref{Dandrea-Book}], accurate
 time synchronization remains a challenging problem especially for turbo-coded systems since they are intended to 
 operate  at very low signal-to-noise ratios (SNRs). In fact, the widespread adoption of turbo codes is in part fueled
 by their ability to operate in the near-Shannon limit even under such adverse SNR conditions [\ref{new_Berrou}]. Yet, the 
 salutary performance of these powerful error-correcting codes is prone to severe degradations if the system is not 
 accurately synchronized in time, phase or frequency. The goal of time synchronization, in particular, consists in
 estimating and compensating for the unknown time delay introduced by the channel so as to provide the decision
 device with symbol-rate samples of lowest possible inter-symbol interference (ISI) corruption  [\ref{Dandrea-Book}].\\
 The problem of timing recovery for linearly-modulated transmissions has been heavily investigated over the last few decades.  A plethora of time delay estimators (TDEs) have been introduced in the open literature and  the vast majority of  existing TDEs are intended to operate with \textit{complete unawareness} of the code structure 
 (see [\ref{vazquez_nda_td}-\ref{Feyh_nda_td}] and references therein). In other words,  the TD estimate is  acquired just  after oversampling the continuous-time signal and then provided to a discrete-time MF in order to output the  symbol-rate  samples. The latter are then used by the turbo decoder, once for all, to  decode the data bits. Therefore, the fact that a  large portion of the data bits is to  become  almost  perfectly known (i.e., correctly decoded) is systematically ignored   by those estimators.   
 The latter are referred to as \textit{non-code-aided} (NCA) or simply NDA TDEs since no \textit{a priori} knowledge about the transmitted symbols is used during the estimation process and, as such, they suffer from severe performance degradations under harsh SNR conditions.  Being more accurate and usually less computationally expensive, DA methods   require, however, the regular transmission of a completely known (i.e., pilot) sequence thereby limiting the whole throughput of the system.\\
%
%
%
%
It sounds reasonable then to conceive a third alternative as a middle ground between these two extreme NDA and DA estimation schemes. Indeed, rather than relying on perfectly known or completely unknown symbols, CA estimation takes advantage of the \textit{soft information} delivered by the decoder at each turbo iteration. In plain English, the decoder assistance is called upon in an attempt to enhance the timing recovery capabilities yet with no impact on the spectral efficiency of the system.
In fact, from one turbo iteration to another, more refined soft information about the conveyed bits are delivered by the two soft-input soft-output (SISO) decoders. These are $i)$ the \textit{a posteriori} LLRs of the code bits and   $ii)$ their extrinsic information. According to the turbo principle, the latter are iteratively exchanged between the two SISO decoders until achieving a steady state whose \textit{a posteriori} LLRs are used as decision metrics  for data detection. In a nutshell, CA  estimation consists simply in leveraging those soft outputs, by embedding the timing recovery task into the decoding process, in an attempt to enhance the estimation performance and vice versa. In the context of timing, phase, and frequency recovery, such CA estimation scheme is usually referred to as \textit{turbo synchronization} [\ref{joint_herzet_noels_invited}].
 A number of CA timing recovery algorithms have been  proposed over the last decade [\ref{LDPC_2013}-\ref{ICC_noels}] and, to the best of the authors' knowledge, only two  approaches are derived from ML theory. 
%
%
 The first one [\ref{icassp_Herzet_TD_Letter}] is based on the well-known expectation maximization (EM) algorithm  whereas the second  [\ref{TSP_Herzet_joint_sum_product_EM}] is a combined sum-product (SP) and EM algorithm approach (i.e., an improvement of [\ref{icassp_Herzet_TD_Letter}]). The SP-EM-based ML estimator offers indeed significant performance improvements over the original EM-based estimator but at the cost of increased computational complexity. In the SP-EM-based ML approach, an EM iteration loop is required in each turbo iteration wherein the algorithm stwitches between the so-called expcetation step (E-STEP) and maximization step (M-STEP). Roughly speaking, in each turbo iteration, the algorithm performs the following main four steps for each EM iteration:
\begin{itemize}
  \item Obtain new symbol-rate samples (via MF) using the TD estimate of the previous EM iteration;
  \item Update the symbols' \textit{a posteriori} probabilities (APoPs) using those new symbol-rate samples;
  \item Marginalize \textit{empirically} the conditional (on the transmitted symbols) likelihood function with respect to those APoPs (E-STEP) ;
  \item Maximize the marginalized LF with respect to the working TD variable (M-STEP).
\end{itemize}

\noindent At the convergence of the EM algorithm,  the obtained TD estimate is used to acquire new ISI-reduced (symbol-rate) samples  which will serve as input for the next turbo iteration where all the aforementioned EM-related steps are repeated.  \\
In this paper, we re-consider the problem of CA time synchronization from both the ``performance bounds'' and ``algorithmic'' point of views. By exploiting the full symmetry of square-QAM constellations and further scrutinizing the Gray coding mechanism, we are able to derive a closed-form and very simple expression for the system's LF. Typically, marginalization of the conditional LF with respect to transmitted symbols is carried-out analytically and the \textit{a priori} LLRs of the elementary code bits are explicitly incorporated  in the LF expression. We propose thereof a more systematic framework to their direct integration in the CA estimation process, thereby  eliminating completely the need for the EM iteration loop  under each turbo iteration. In other words,  the new LF needs to be maximized only once per-turbo iteration (contrarily to SP-EM) after being updated by the associated \textit{a priori} LLRs which are computed from the output of the SISO decoders. Consequently, the proposed CA timing recovery algorithm offers significant improvements in computational complexity as compared to the existing SP-EM. As a matter of fact, the new algorithm is 35 and 70 times less computationally complex than SP-EM for 64 and 256 QAMs, respectively. It also enjoys an advantage in terms of estimation accuracy for low SNR levels and higher-order modulations. \\  
From the ``performance bounds'' point of view, we also tackle the analytical derivation of the \textit{stochastic} CRLBs for the underlying CA estimation problem. Actually, unlike many other loose bounds, the \textit{stochastic}\footnote{In linearly-modulated transmissions, the  \textit{stochastic} model refers to estimation under the assumption of unknown and \textit{random} transmitted symbols. This to be opposed to the \textit{deterministic} model wherein the symbols are assumed to be unknown but \textit{not random} [\ref{vazquez_nda_td}].} CRLB is a fundamental lower bound that reflects the actual achievable performance in practice [\ref{Kay-book}].
Yet, even under uncoded transmissions, the complex structure of the LF makes it extremely hard, if not impossible, to derive analytical expressions for this practical bound, especially with high-order modulations. Therefore, in the specific context of timing recovery, the stochastic CRLBs were previously evaluated  using exhaustive Monte-Carlo simulations (i.e., \textit{empirically}) in  [\ref{noels_TD_NDA}] and [\ref{EURASIP_noels_1}] for both NCA and CA estimations, respectively. Just recently though were their analytical expressions established [\ref{masmoudi_CRLB_NDA}] but only in the NCA (i.e., NDA) case.\\
In this paper, we succeed in factorizing the LF of the coded system  as the product of two analogous terms involving two random variables that are \textit{almost} identically distributed, i.e.,  their probability density functions (pdfs) have the same expression but parametrized differently. We then capitalize on this interesting property to derive, for the very first time, the closed-form expressions for the  TD CA CRLBs from arbitrary turbo-coded square-QAM-modulated transmissions.  The new closed-form expressions corroborate the previous attempts reported in [\ref{EURASIP_noels_1}] to evaluate the TD CA CRLBs \textit{empirically}  and offer a way to their immediate evaluation in practice. Moreover, as will be shown later, the previously published closed-form NDA CRLBs [\ref{masmoudi_CRLB_NDA}] boil down to a very special case of the new closed-form CA CRLBs  by simply setting all the code bits'   \textit{a priori} LLRs to zero. \\
The rest of this paper is structured as follows. In section II, we present the system model. 
 In section III, we derive the expression of the log-likelihood function (LLF)  and express it explicitly as function of the coded bits' \textit{a priori} LLRs. in Section IV,  we establish the new closed-form expressions for the TD CA CRLBs. In Section V, we introduce the new CA ML time delay estimator. In Section VI, we discuss the simulation results of the proposed CA ML estimator and closed-form CRLBs. Finally, we draw out some concluding remarks in section VII.\\
We also mention beforehand that some of the common notations will be used in this paper.  Vectors and matrices are represented in lower- and upper-case bold fonts, respectively. $\mathbf{I}_N$ and $\mathbf{0}_N$ denote, respectively, the $N\times N$ identity matrix and the $N-$dimensional all-zero  vector. The shorthand notation $\mathbf{x}\sim\mathcal{N}(\mathbf{m},\mathbf{R})$ means that the vector $\mathbf{x}$ follows a normal (i.e., Gaussian) distribution with mean $\mathbf{m}$ and auto-covariance matrix $\mathbf{R}$. Moreover, $\{.\}^T$ and $\{.\}^H$ denote the transpose and the Hermitian (transpose conjugate) operators, respectively. The operators $\Re\{.\}$ and $\Im\{.\}$ return, respectively, the real and imaginary parts  of any complex number. The operators $\{.\}^*$ and $|.|$ return its conjugate and its amplitude, respectively, and  $j$ is the pure complex number that verifies $j^{2}=-1$. The Kronecker and Dirac delta functions are denoted, respectively, as  $\delta_{m,n}$ and $\delta(t)$. We will also denote the probability mass function (PMF) for  discrete random variables (RVs) by $P[.]$ and the pdf for continuous RVs  by $p[.]$ The statistical expectation is denoted as $\mathbb{E}\{.\}$ and the notation $\triangleq$ is used for definitions.        
\section{System Model}\label{section_2}
Consider a  turbo-coded system where a binary sequence of information bits is fed into a turbo encoder consisting of two identical recursive and systematic convolutional codes (RSCs) which are concatenated in parallel via an inner interleaver $\Pi_1$. The resulting code bits are fed into a puncturer which selects an appropriate combination of the parity bits, from both encoders, in order to achieve a desired code rate $R$. The entire code bit sequence is then scrambled with an outer interleaver, $\Pi_2$, and divided into $K$ subgroups of $2p$ bits each for some integer $p\geq 1$. The $k^{th}$ subgroup of code bits, $b_1^{k}b_2^{k}\cdots b_l^{k}\cdots b_{2p}^{k}$, is conveyed by a symbol $a(k)$ that is selected from a fixed alphabet, $\mathcal{C}_p=\{c_0, c_1,\cdots, c_{M-1}\}$, of a  $M-$ary (with $M=2^{2p}$) QAM constellation (i.e., square-QAM). In fact, each point, $c_m\in \mathcal{C}_p$, is mapped onto a unique sequence of $\log_2(M)=2p$ bits denoted here as $ \bar{b}_1^{m}\bar{b}_2^{m}\cdots \bar{b}_l^{m}\cdots \bar{b}_{2p}^{m}$, according to the Gray coding mechanism, and  the point $c_m$ is selected to convey the $k^{th}$  code bits subgroup [i.e., $a(k)=c_m$] if and only if $b_l^{k}=\bar{b}_l^{m}$ for $l=1,2,\cdots,2p$.
We also  define the \textit{a priori} LLR of the $l^{th}$  code bit, $b^{k}_{l}$, conveyed by $a(k)$ as follows:
\begin{eqnarray}\label{E13}
L_{l}(k)~\triangleq~\ln\left(\frac{P[b^{k}_{l}=1]}{{P[b^{k}_{l}=0]}}\right).
\end{eqnarray}
Using  (\ref{E13}) and the fact that $P[b_l^k=0]+P[b_l^k=1]=1$, it can be easily shown that:
\begin{eqnarray}\label{apriori_1}
P[b_l^k=1]= \displaystyle \frac{e^{L_l(k)}}{1+e^{L_l(k)}}~~~~\textrm{and} ~~~~ P[b_l^k=0]=\displaystyle \frac{1}{1+e^{L_l(k)}}.
\end{eqnarray}
For mathematical convenience,  the two identities in (\ref{apriori_1}) can be merged together to yield  the following common generic expression:
\begin{eqnarray}\label{bit_prob_generic}
P[b_l^{k}=\bar{b}_l^{m}]\!=\!\frac{1}{2\cosh\big(L_{l}(k)/2\big)}\mathlarger{e^{(\bar{b}_l^{m}-1)\frac{L_{l}(k)}{2}}},
\end{eqnarray}
in which $\bar{b}_l^{m}$ is either $0$ or $1$ depending on which of the symbols $c_m$ is transmitted, at time instant $k$, and of course on the Gray mapping associated to the constellation.
%
%
The obtained information-bearing symbols, $\{a(k)\}_{k=1}^{K}$, are then pulse-shaped and the resulting continuous-time  signal:
\begin{equation}\label{Output}
x(t)=\sum_{k=1}^{K}a(k)~\!h(t-kT),
\end{equation}
 is transmitted over the communication channel with $T$ being the symbol duration and $h(t)$ a unit-energy square-root shaping pulse. Being completely unknown to the receiver \textit{a priori}, the transmitted symbols $\{a(k)\}_k$   are drawn from a given $M$-ary Gray-coded (GC) square-QAM constellation whose alphabet is denoted as $\mathcal{C}_p=\{c_0, c_1,\cdots, c_{M-1}\}$. Here, by \textit{square} QAM we mean $M=2^{2p}$ for some integer $p\geq 1$ \big(i.e., QPSK, $16-$QAM, $64-$QAM, etc...\big).   The Nyquist pulse $g(t)$ obtained from $h(t)$ is defined as:
\begin{equation}\label{nyquist}
g(t) = \int_{-\infty}^{+\infty} h(x)h(t+x)dx,
\end{equation}
and satisfies the first Nyquist criterion [\ref{Dandrea-Book}]:
\begin{equation}\label{nyquist_condition}
g(nT) = 0,~\textrm{for any integer}~n \neq 0.
\end{equation}
At the receiver side, assuming perfect frequency and phase synchronizations, the (delayed)  continuous-time received signal before matched filtering is expressed as:
\begin{equation}\label{Input-Output-relationship}
y(t)=\sqrt{E_s}~x(t-\tau)~+~w(t),
\end{equation}
where $E_s$ is the transmit signal energy and $\tau$ is the unknown time delay parameter to be estimated. Moreover, $w(t)$ is a \textit{proper} complex additive white Gaussian noise (AWGN) with independent real and imaginary parts, each of variance $\sigma^2$ (i.e., with overall noise power $N_0=2\sigma^2$).
The SNR of the channel is also denoted as:
\begin{eqnarray}
\rho~\triangleq~ \frac{E_s}{N_0}~=~\frac{E_s}{2\sigma^2}.
\end{eqnarray}
An integral step in the derivation of \textit{stochastic} ML estimators and  CRLBs consists in finding the LLF of the system. This requires marliginalizing the conditional (on the unknown  symbols) LF over the constellation alphabet.  In \textit{completely} NDA estimation (or before data detection), no \textit{a priori} information  is available about the transmitted symbols. Therefore, the latter are usually assumed to be equally likely, i.e., with equal \textit{a priori} probabilities (APPs). That is to say $\forall~\!c_m\in\mathcal{C}_p$:
\begin{equation}
P[a(k)=c_m]=\frac{1}{M}~~~~~~ \textrm{for}~~ k=1,2,\cdots,K.
\end{equation}
%
  In CA estimation, however, the actual APPs of the transmitted symbols must be used in order to enhance the estimation performance as  done in the next section. By doing so, we will ultimately express the LLF explicitly as function of the \textit{a priori} LLRs of the individual coded bits.
%
 As will be explained later in Section \ref{CRLB_evaluation}, accurate estimates  for the underlying  LLRs  can be obtained in practice from the soft outputs of the two SISO decoders at the convergence of the BCJR algorithm \cite{BCJR}.\\

\section{Derivation of the LLF}\label{section_3}
As widely known, the set of finite-energy signals usually denoted as:
\begin{eqnarray}
\mathcal{L}^2_{\mathbb{R}}=\left\{s(t)~\textrm{such that}~\int_{\mathbb{R}}|s(t)|^2dt<+\infty\right\}, \nonumber
\end{eqnarray}
form an infinite-dimension Hilbert subspace [\ref{Mallat}] that can be endowed with an orthonormal basis $\{\varphi_n(t)\}_{n}$ and an inner product as follows:
\begin{eqnarray}
\langle s_1(t),s_2(t)\rangle=\int_{\mathbb{R}}s_1(t)s_2(t)^*dt, ~~\forall~s_1(t),s_2(t)\in \mathcal{L}^2_{\mathbb{R}}.
\end{eqnarray}
Therefore, an exact discrete representation for any continuous-time signal $s(t)\in \mathcal{L}^2_{\mathbb{R}}$ requires an infinite-dimensional vector, ${\bm s}$, that contains its expansion coefficients, $\big\{s_n=\langle s(t),\varphi_n(t)\rangle\big\}_n$, in the basis $\{\varphi_n(t)\}_{n}$.   To sidestep this problem, we first consider the $N$-dimensional truncated representation vectors:
\begin{eqnarray}
\bm{y}_N&\!\!\!\!=\!\!\!\!&\left[y_1,y_2,\ldots,y_N\right]^T\!\!,\\
\bm{w}_N&\!\!\!\!=\!\!\!\!&\left[w_1,w_2,\ldots,w_N\right]^T\!\!,\\
\bm{x}_N(\tau)&\!\!\!\!=\!\!\!\!&\left[x_1(\tau),x_2(\tau),\ldots,x_N(\tau)\right]^T\!\!.
\end{eqnarray}
 that contain  the orthogonal projection coefficients  of $y(t)$, $w(t)$, and $x(t-\tau)$, respectively, over the first $N$ basis functions $\{\varphi_n(t)\}_{n=1}^N$ (for any $N\geq 1$), i.e.:
\begin{eqnarray}
\label{projection_y}y_n&\!\!\!\!=\!\!\!\!&\big\langle y(t),\varphi_n(t)\big\rangle=\int_{\mathbb{R}}y(t)\varphi_n(t)^*dt,\\
\label{projection_w}w_n&\!\!\!\!=\!\!\!\!&\big\langle w(t),\varphi_n(t)\big\rangle=\int_{\mathbb{R}}w(t)\varphi_n(t)^*dt,\\
\label{projection_x}x_n(\tau)&\!\!\!\!=\!\!\!\!&\big\langle x(t-\tau),\varphi_n(t)\big\rangle=\int_{\mathbb{R}}x(t-\tau)\varphi_n(t)^*dt,
\end{eqnarray}
Using (\ref{Input-Output-relationship}) and (\ref{projection_y}) to (\ref{projection_x}), it follows that:
\begin{eqnarray}\label{input-output-vector}
\bm{y}_N=\sqrt{E_s}~\!\bm{x}_N(\tau)+\bm{w}_N.
\end{eqnarray}
Due to the orthogonality of the basis functions, it can be shown that the noise projection coefficients, $\{w_n\}_{n=1}^N$, explicitly given by (\ref{projection_w})   are uncorrelated, i.e,  $\mathbb{E}\big\{w_nw_m^*\big\}=2\sigma^2\delta_{n,m}$.
Hence, they are independent since they are also Gaussian-distributed\footnote{This is because they are obtained by some linear transformations (i.e., the orthogonal projection) of the original continuous-time white Gaussian random process $w(t)$.}
 leading to $\bm{w}_N\sim\mathcal{N}(\mathbf{0}_N,2\sigma^2\mathbf{I}_N)$.
Therefore, the pdf of the vector  ${\bm y}_N$ in (\ref{input-output-vector}) conditioned on the sequence of transmitted symbols, $\bm a=[a(1), a(2),\ldots,a(K)]^T$, and parametrized by $\tau$ is given by:
\begin{eqnarray}\label{truncated_pdf}
\!\!\!\!\!\!\!\!p({\bm{y}}_N|\bm a;\tau)&=&\prod\displaylimits_{n=1}^N\mathsmaller{\frac{1}{2\pi\sigma^2}}\exp\!\left\{\!-\mathsmaller{\frac{1}{2\sigma^2}}\big|y_n\!-\!\mathsmaller{\mathsmaller{\sqrt{E_s}}}x_n(\tau)\big|^2\right\}.
\end{eqnarray}
Note here that, although we do not show it explicitly, the transmitted symbols are indeed involved in (\ref{truncated_pdf}) via the coefficients $\{x_n(\tau)\}_{n}$.  After dropping the constant terms that do not depend explicitly on $\tau$ in (\ref{truncated_pdf}), we obtain the simplified \textit{truncated} LF: 
\begin{eqnarray}\label{truncated_LF}
\!\!\!\!\!\!\!\!\!\!\!\!\Lambda({\bm {y}}_N|\bm a;\tau)&=&\exp\!\left\{\!\mathsmaller{\frac{\sqrt{E_s}}{\sigma^2}}\sum\displaylimits_{n=1}^N\!\Re\left\{y_n x_n(\tau)^*\right\}-\mathsmaller{\frac{E_s}{2\sigma^2}}\!\sum_{n=1}^N\!\big|x_n(\tau)\big|^2\!\right\}.
\end{eqnarray}
The conditional LF which incorporates all the information  contained in the \textit{non-truncated} vector ${\bm y}$ \big[or equivalently  the received continuous-time  signal $y(t)$\big], is obtained by  making $N$ tend to infinity in (\ref{truncated_LF}). By doing so and using  the Plancherel equality, we obtain the following  \textit{conditional} LF:
\begin{eqnarray}\label{conditional_LLF_2_nonsimplified}
\!\!\!\!\!\!\!\!\!\!\Lambda(\bm {y}|\bm a;\tau)&=&\textstyle\exp\!\left\{\!\frac{\sqrt{E_s}}{\sigma^2}\!\mathlarger{\int_{\mathbb{R}}}\!\Re\big\{ y(t) x(t\!-\!\tau)^*\big\}dt-\textstyle\frac{E_s}{2\sigma^2}\mathlarger{\int_{\mathbb{R}}}|x(t\!-\!\tau)|^2dt\!\right\}.
\end{eqnarray}
Now, replacing the transmitted signal $x(t)$ by its expression given in (\ref{Output}), and exploiting the fact that the  shaping pulse, $g(t)$, in (\ref{nyquist}) verifies the first-order Nyquist criterion (\ref{nyquist_condition}), it can be shown that:
\begin{eqnarray}\label{conditional_LLF_3}
\Lambda(\bm {y}|\bm a;\tau)&\!\!\!\!=\!\!\!\!&\prod_{k=1}^{K}\Omega_{\tau}\big(a(k),y(t)\big),
\end{eqnarray}
where
%
\begin{eqnarray}\label{omega_nonaveraged}
\!\!\!\!\!\!\!\!\!\!\!\!\Omega_{\tau}\big(a(k),y(t)\big)&\triangleq&\exp\!\left\{\!\mathsmaller{\frac{\sqrt{E_s}}{\sigma^2}}\!\!\!\int_{\mathbb{R}}\!\!\Re\big\{ y(t) a(k)^*\!\big\}h(t\!-\!kT\!-\!\tau)dt\!-\!\mathsmaller{\frac{E_s}{2\sigma^2}}\big|a(k)\big|^2\right\}.
\end{eqnarray}
The \textit{unconditional} LF, $\Lambda(\bm {y};\tau)$, is obtained by averaging (\ref{conditional_LLF_3}) over all  possible  transmitted symbol sequences of size $K$, i.e., $\Lambda(\bm {y};\tau)=\mathbb{E}_{\bm{a}}\{\Lambda(\bm {y}|\bm a;\tau)\}$ leading to:
\begin{eqnarray}\label{unconditional_LLF}
\Lambda(\bm {y};\tau)&\!\!\!\!=\!\!\!\!&\sum_{\bm{c}_i\in\mathcal{C}_p^K}P[\bm{a}=\bm{c}_i]\Lambda(\bm {y}|\bm a=\bm{c}_i;\tau).
\end{eqnarray}
Under coded digital transmissions, a simplifying assumption is usually used in  estimation practices, whether CA or NCA, in order to allow for tractable mathematical derivations of  CRLBs and  ML estimators of any parameter. This assumption postulates that the transmitted symbols are  independent  (cf. [\ref{LDPC_2013}-\ref{noels_TD_NDA}]   and references therein) in spite of the statistical dependence between the coded bits that is introduced  by channel coding.  In fact, before even initiating  the decoding process itself, the system needs  to be fully synchronized by estimating the time delay, as well as, the phase and frequency offsets. Moreover, the decoder itself needs some estimates for other key channel parameters, e.g., the channel coefficient, noise variance, SNR, etc. All those estimates  are obtained by applying traditional NDA estimators directly on the  symbol-rate  samples that are delivered by the matched filter before starting data decoding.  As a matter of fact, in digital transmissions, all state-of-the-art NDA estimators (for any parameter, whether maximum likelihood or moment-based) are indeed based on the assumption of independent symbols although the latter are actually dependent due to channel coding.\\
 We emphasize, however, that exploitation of this assumption does not imply denying to exploit the dependence of the coded bits during the decoding process itself.  Indeed, such dependence is exploited by the SISO decoders in order to output the estimates for the coded bits' \textit{a posteriori} LLRs. The latter are then used to decode the bits and also to compute their \textit{a priori} LLRs (as explained later in Section \ref{CRLB_evaluation}) which are in turn used to evaluate the CA CRLBs and to find the CA TD ML estimate.
 Yet, even by assuming independent symbols (both in this paper and all existing works), it turns out that no much information is lost from the estimation point of view. In fact, the resulting CA estimation schemes achieve the ideal data-aided one (where all the symbols are perfectly known) over a wide range of practical SNRs where the  completely NDA schemes do not (cf. Figs. \ref{estimator-QPSK} and \ref{estimator-16-QAM} in this paper and the reported simulation results in other researchers' works).  
%
%
%
 Using the assumption of independent symbols  it follows that:
%
%
%
\begin{eqnarray}\label{independent_symbols}
P\big[\bm{a}=\bm{c}_i\big]=\prod_{k=1}^{K}P\big[a(k)=\bm{c}_i(k)\big].
\end{eqnarray}
Plugging (\ref{conditional_LLF_3}) and (\ref{independent_symbols}) in (\ref{unconditional_LLF}), it can be shown that:
\begin{eqnarray}
\!\!\!\!\!\!\!\!\Lambda(\bm {y};\tau)&\!\!\!\!=\!\!\!\!&\sum_{\bm{c}_i\in\mathcal{C}_p^K}\prod_{k=1}^{K}P\big[a(k)=\bm{c}_i(k)\big]\Omega_{\tau}\big(\bm{c}_i(k),y(t)\big)\nonumber\\
\!\!\!\!\!\!\!\!&\!\!\!\!=\!\!\!\!&\prod_{k=1}^{K}\sum_{c_m\in\mathcal{C}_p}P\big[a(k)=c_m\big]\Omega_{\tau}\big(c_m,y(t)\big).
\end{eqnarray} 
Therefore, the \textit{unconditional} log-likelihood function (LLF) defined as $\mathcal{L}(\bm {y};\tau)\triangleq\ln\big(\Lambda(\bm {y};\tau)\big)$, is given by:
\begin{eqnarray}\label{LLF}
\mathcal{L}(\bm {y};\tau)&\!\!\!\!=\!\!\!\!&\sum_{k=1}^{K}\ln\Big(\bar{\Omega}_k\big(\tau,y(t)\big)\Big),
\end{eqnarray}
in which $\bar{\Omega}_k\big(\tau,y(t)\big)$ is simply the average of $\Omega_{\tau}\big(a(k),y(t)\big)$ over  the constellation alphabet, i.e.:
\begin{eqnarray}\label{H_definition}
\bar{\Omega}_k\big(\tau,y(t)\big)&\!\!\!\!\triangleq\!\!\!\!&\sum_{c_m\in\mathcal{C}_p}P\big[a(k)=c_m\big]\Omega_{\tau}\big(c_m,y(t)\big).
\end{eqnarray}
For ease of notations, we will hereafter no longer show the dependence of $\bar{\Omega}_k\big(\tau,y(t)\big)$ on the received signal, $y(t)$, and denote it simply as  $\bar{\Omega}_k(\tau)$. Next, we will further manipulate this term  and ultimately factorize it into two analogous terms which involve two independent and \textit{almost} identically distributed RVs. In fact, by further  denoting the top-right quadrant of the constellation as $\widetilde{\mathcal{C}}_p$, it follows that $\mathcal{C}_p=\widetilde{\mathcal{C}}_p\cup(-\widetilde{\mathcal{C}}_p)\cup\widetilde{\mathcal{C}}_p^*\cup(-\widetilde{\mathcal{C}}_p^*)$. Thus, the sum over $c_m\in\mathcal{C}_p$ in (\ref{H_definition}) can be equivalently replaced by a sum over each $\tilde{c}_m\in\widetilde{\mathcal{C}}_p$ and its three symmetrical points in the other quadrants. By doing so  and noticing that $|\tilde{c}_m|\!=\!|\!-\!\tilde{c}_m|\!=\!|\tilde{c}^*_m|\!=\!|\!-\!\tilde{c}^*_m|$, we obtain from (\ref{omega_nonaveraged}) and (\ref{H_definition}):   
\begin{eqnarray}\label{H_expression_1}
\!\!\!\bar{\Omega}_k\big(\tau\big)&=&\!\!\!\sum_{\tilde{c}_m\in \widetilde{\mathcal{C}}_p} e^{-\frac{{E_s}}{2\sigma^2}|\tilde{c}_m|^2}\!\times\Bigg(\textstyle\! P\big[a(k)\!=\!\tilde{c}_m\big]\exp\bigg\{\frac{\sqrt{E_s}}{\sigma^2}\mathlarger{\int_{\mathbb{R}}}\!\Re\{\tilde{c}^*_m y(t)\}h(t\!-\!kT\!-\!\tau)dt\bigg\}\nonumber\\
\!\!\!&\!\!\!\!\!\!\!\!&~~~~~~~~~~+\textstyle P\big[a(k)\!=\!-\tilde{c}_m\big]\exp\bigg\{\frac{\sqrt{E_s}}{\sigma^2}\mathlarger{\int_{\mathbb{R}}}\!\Re\{-\tilde{c}^*_m y(t)\}h(t\!-\!kT\!-\!\tau)dt\bigg\}\nonumber\\
 \!\!\!&\!\!\!\!\!\!\!\!&~~~~~~~~~~+\textstyle P\big[a(k)\!=\!\tilde{c}^*_m\big]\exp\bigg\{\frac{\sqrt{E_s}}{\sigma^2}\mathlarger{\int_{\mathbb{R}}}\!\Re\{\tilde{c}_m y(t) \}h(t\!-\!kT\!-\!\tau)dt\bigg\}\nonumber\\
 \!\!\!&\!\!\!\!\!\!\!\!&~~~~~~~~~~+\textstyle P\big[a(k)\!=\!-\tilde{c}^*_m\big]\exp\bigg\{\frac{\sqrt{E_s}}{\sigma^2}\mathlarger{\int_{\mathbb{R}}}\!\Re\{-\tilde{c}_m y(t)\}h(t\!-\!kT\!-\!\tau)dt\bigg\}\!\Bigg).
\end{eqnarray}
Using a simple recursive scheme that allows the construction of arbitrary square-QAM constellations, it has been recently shown in [\ref{Bellili-CA-SNR}] that the APPs for each symbol $x(k)$ are expressed as follows ($\forall~\!\tilde{c}_m\in\widetilde{\mathcal{C}}_p$):
\begin{eqnarray}\label{prob_LSB_explicit_1}
\!\!\!\!\!\!\!\!\!P[x(k)\!=\!\tilde{c}_m]&\!\!\!\!\!=\!\!\!\!\!&\beta_{k}~\mu_{k,p}(\tilde{c}_{m})~e^{\frac{L_{2p-1}(k)}{2}}e^{\frac{L_{2p}(k)}{2}},\\
\!\!\!\!\!\!\!\!\!P[x(k)\!=\!\tilde{c}_m^*]&\!\!\!\!\!=\!\!\!\!\!&\beta_{k}~\mu_{k,p}(\tilde{c}_{m})~e^{-\frac{L_{2p-1}(k)}{2}}e^{\frac{L_{2p}(k)}{2}},\label{prob_LSB_explicit_2}\\
\!\!\!\!\!\!\!\!\!P[x(k)\!=\!-\tilde{c}_m]&\!\!\!\!\!=\!\!\!\!\!&\beta_{k}~\mu_{k,p}(\tilde{c}_{m})~e^{-\frac{L_{2p-1}(k)}{2}}e^{-\frac{L_{2p}(k)}{2}},\label{prob_LSB_explicit_3}\\
\!\!\!\!\!\!\!\!\!P[x(k)\!=\!-\tilde{c}_m^*]&\!\!\!\!\!=\!\!\!\!\!&\beta_{k}~\mu_{k,p}(\tilde{c}_{m})~e^{\frac{L_{2p-1}(k)}{2}}e^{-\frac{L_{2p}(k)}{2}},\label{prob_LSB_explicit_4}
\end{eqnarray}
in which $\mu_{k,p}(\tilde{c}_{m})$ and $\beta_{k}$ are given by:
\begin{eqnarray}\label{mu_c_m}
\mu_{k,p}(\tilde{c}_{m})&\triangleq\displaystyle&\prod_{l=1}^{2p-2}\mathlarger{e^{(2\bar{b}_l^{m}-1)\frac{L_{l}(k)}{2}}}, ~~~~\forall~\tilde{c}_m\in\widetilde{\mathcal{C}}_p\\
\beta_{k}&\triangleq\displaystyle&\displaystyle\prod_{l=1}^{2p}\frac{1}{2\cosh\big(L_{l}(k)/2\big)}.
\end{eqnarray}
Plugging  (\ref{prob_LSB_explicit_1})-(\ref{prob_LSB_explicit_4}) back into (\ref{H_expression_1}) and using the trivial identity $e^x+e^{-x}=2\cosh(x)$, it can be shown that:
\begin{eqnarray}\label{H_expression_2}
\!\!\!\!\bar{\Omega}_k(\tau)&\!\!\!\!=\!\!\!\!&2\beta_{k}\!\!\sum_{\tilde{c}_m\in \widetilde{\mathcal{C}}_p} \mu_{k,p}(\tilde{c}_m)\mathlarger{e^{-\rho|\tilde{c}_m|^2}}\!\times\Bigg[\textstyle\! \cosh\!\bigg\{\!\frac{\sqrt{E_s}}{\sigma^2}\!\mathlarger{\int_{\mathbb{R}}}\!\Re\{\tilde{c}_m y(t)\}h(t\!-\!kT\!-\!\tau)dt+\frac{L_{2p}(k)-L_{2p-1}(k)}{2}\!\bigg\}+\nonumber\\
\!\!\!\!&\!\!\!\!\!\!\!\!&~~~~~~~~~~~~~~~~~~~~~~~~~~~~~~\textstyle\cosh\!\bigg\{\!\frac{\sqrt{E_s}}{\sigma^2}\!\mathlarger{\int_{\mathbb{R}}}\!\Re\{\tilde{c}_m^* y(t) \}h(t\!-\!kT\!-\!\tau)dt+\frac{L_{2p}(k)+L_{2p-1}(k)}{2}\!\bigg\}\Bigg]\!.
\end{eqnarray}
Furthermore, by using the relationship $ \cosh(x)+\cosh(y)= 2 \cosh(\frac{x+y}{2}) \cosh(\frac{x-y}{2})$ along with the two identities $\tilde{c}_m+\tilde{c}_m^*=2\Re\{\tilde{c}_m\}$ and $\tilde{c}_m-\tilde{c}_m^*=2j\Im\{\tilde{c}_m\}$, it can be shown that (\ref{H_expression_2}) can be rewritten as follows:
\begin{eqnarray}\label{H_expression_3}
\!\!\!\!\bar{\Omega}_k(\tau)&\!\!\!\!=\!\!\!\!&4\beta_{k}\!\!\!\!\sum_{\tilde{c}_m\in \widetilde{\mathcal{C}}_p} \!\!\!\textstyle\mu_{k,p}(\tilde{c}_m)\mathlarger{e^{-\rho|\tilde{c}_m|^2}}\!\textstyle \cosh\!\bigg\{\!\frac{\sqrt{E_s}\Re\{\tilde{c}_m\}}{\sigma^2}u_k(\tau)\!+\!\frac{L_{2p}(k)}{2}\!\bigg\}\times\textstyle\cosh\bigg\{\!\frac{\sqrt{E_s}\Im\{\tilde{c}_m\}}{\sigma^2}v_k(\tau)+\frac{L_{2p-1}(k)}{2}\!\bigg\}\Bigg],
\end{eqnarray}
in which $u_k(\tau)$ and $v_k(\tau)$ are the matched-filtered \textit{in-phase} and \textit{quadrature} components of the received signal given by:
\begin{eqnarray}
\label{inphase}u_k(\tau)&\!\!\!\!=\!\!\!\!&\int_{-\infty}^{+\infty}\!\Re\big\{y(t)\big\}h(t-kT-\tau)dt,\\
\label{quadrature}v_k(\tau)&\!\!\!\!=\!\!\!\!&\int_{-\infty}^{+\infty}\!\Im\big\{y(t)\big\}h(t-kT-\tau)dt.
\end{eqnarray}
 Since in the Cartesian coordinate system of the constellation each $\tilde{c}_m\in\widetilde{\mathcal{C}}_p$  can be written\footnote{Note here that $d_p$ is half the minimum inter-symbol distance whose expression is given in [\ref{Bellili-CA-SNR}, eq. (30)]  explicitly as function of $p$  for normalized-energy constellations.} as  $\tilde{c}_m= [2i-1]d_{p}+ j[2n-1]d_{p}$ for some $1\leq i,n\leq 2^{p-1}$, then  the single sum over $\tilde{c}_m$ in (\ref{H_expression_3}) can be equivalently replaced by a double sum over the two counters $i$ and $n$ as follows: 
\begin{eqnarray}\label{H_expression_4}
\!\!\!\!\!\!\!\!\!\bar{\Omega}_k(\tau)&\!\!\!\!=\!\!\!\!&4\beta_{k}\sum_{i=1}^{2^{p-1}}\!\sum_{n=1}^{2^{p-1}}\Bigg[\mu_{k,p}\Big([2i-1]d_{p}+ j[2n-1]d_{p}\Big)\nonumber\\
\!\!\!\!\!\!\!\!\!&\!\!\!\!\!\!\!\!&~~~~~~~~~~\times~\!\mathlarger{e^{-\rho[2i-1]^2d_p^2}}
\cosh\bigg(\frac{\sqrt{E_s}[2i-1]d_p}{\sigma^2}u_k(\tau)+\frac{L_{2p}(k)}{2}\bigg)\nonumber\\
\!\!\!\!\!\!\!\!\!&\!\!\!\!\!\!\!\!&~~~~~~~~~~\times~\!\mathlarger{e^{-\rho[2n-1]^2d_p^2}}\cosh\!\bigg(\!\!\frac{\sqrt{E_s}[2n-1]d_p}{\sigma^2}v_k(\tau)\!+\!\frac{L_{2p-1}(k)}{2}\!\bigg)\Bigg]\!.
\end{eqnarray}
We also recall the  following decomposition recently shown in [\ref{Bellili-CA-SNR}] for each $\tilde{c}_{m}=[2i-1]d_{p}+ j[2n-1]d_{p}\in\widetilde{\mathcal{C}}_p$:  
\begin{eqnarray}\label{mu_c_m_factorized}
\mu_{k,p}\big([2i-1]d_{p}+ j[2n-1]d_{p}\big)=\theta_{k,2p}(i)\theta_{k,2p-1}(n)~~~~,
\end{eqnarray}
where 
\begin{eqnarray}\label{theta_2p}
\theta_{k,2p}(i)&\!\triangleq\!&\displaystyle\prod_{l=1}^{p-1}\mathlarger{e^{(2\bar{b}_{2l}^{(i)}-1)\frac{L_{2l}(k)}{2}}},\\
\label{theta_2p-1}\theta_{k,2p-1}(n)&\!\triangleq\!&\displaystyle\prod_{l=1}^{p-1}\mathlarger{e^{(2\bar{b}_{2l-1}^{(n)}-1)\frac{L_{2l-1}(k)}{2}}}.
\end{eqnarray}
After using (\ref{mu_c_m_factorized}) in (\ref{H_expression_4}) and splitting the two sums, we obtain the following much useful factorization for $\bar{\Omega}_k(\tau)$:
\begin{equation}\label{h_tau=F*F}
\bar{\Omega}_k(\tau)=4\beta_{k}~\!\!F_{k,2p}\big(u_k(\tau)\big)~\!\!F_{k,2p-1}\big(v_k(\tau)\big),
\end{equation}
where
\begin{eqnarray}\label{function_F}
\!\!\!\!\!\!\!\!\!\!\!F_{k,q}(x)&=&\!\!\sum_{i=1}^{2^{p-1}}\!\textstyle\! \theta_{k,q}(i)\mathlarger{e^{-\rho[2i-1]^2d_p^2}}\cosh\!\bigg(\!\!\frac{\sqrt{E_s}[2i-1]d_p}{\sigma^2}x+\frac{L_{q}(k)}{2}\!\!\bigg),
\end{eqnarray}
in which $q$ is a generic counter that is used from now on to refer to $2p$ or $2p-1$ depending on the context.
Finally, by using (\ref{h_tau=F*F}) back in (\ref{LLF}) and dropping the constant term $4\beta_{k}$ that do not depend on $\tau$, the useful LLF develops into:
\begin{eqnarray}\label{Final_LLF}
\mathcal{L}(\bm {y};\tau)&=&\sum_{k=1}^{K}\ln\!\Big(\!F_{k,2p}\big(u_k(\tau)\big)\!\Big)~+~\sum_{k=1}^{K}\ln\!\Big(\!F_{k,2p-1}\big(v_k(\tau)\big)\!\Big).
\end{eqnarray}
We succeeded here in decomposing the LLF into two analogous terms [the two sums in (\ref{Final_LLF})] involving each either RVs $u_k(\tau)$ or $u_k(\tau)$ that will be shortly shown to have almost the same distributions. This is  actually the cornerstone result upon which we will establish  the analytical expressions for the CA TD CRLBs  in the next section.
\section{Derivation of the CA CRLBs}
As an overall benchmark, the CRLB lower bounds the variance of any unbiased estimator,  $\widehat{\tau}$, of the time delay parameter, i.e., $\mathbb{E}\big\{(\widehat{\tau}-\tau)^{2}\big\}\geq\textrm{CRLB}(\tau)$. It is explicitly given by [\ref{Kay-book}]:
\begin{eqnarray}
\textrm{CRLB}(\tau)=\frac{1}{I(\tau)},
\end{eqnarray}   
where $I(\tau)$ is the so-called Fisher information for the received data which is given by:
\begin{eqnarray}\label{Fisher_dinformation_definition}
I(\tau)=-\mathbb{E}\left\{\frac{\partial^2\mathcal{L}(\bm {y};\tau)}{\partial\tau^2}\right\}.
\end{eqnarray}
Using (\ref{Final_LLF}) in (\ref{Fisher_dinformation_definition}) and owing to the linearity of the partial derivative and expectation operators, it immediately follows that:
\begin{eqnarray}\label{Fisher_dinformation_1}
I(\tau)&\!\!\!\!=\!\!\!\!&\sum_{k=1}^{K}\big[\gamma_{k,2p}(\tau)+\gamma_{k,2p-1}(\tau)\big],
\end{eqnarray}
where
\begin{eqnarray}
\label{gamma_k_2p}\gamma_{k,2p}(\tau)&\!\!\!\triangleq\!\!\!&-\mathbb{E}\left\{\partial^2 \ln\big(F_{k,2p}\big(u_k(\tau)\big)\big)/\partial\tau^2\right\},\\
\label{gamma_k_2p-1}\gamma_{k,2p-1}(\tau)&\!\!\!\triangleq\!\!\!&-\mathbb{E}\left\{\partial^2 \ln\big(F_{k,2p-1}\big(v_k(\tau)\big)\big)/\partial\tau^2\right\}.
\end{eqnarray}
 Before delving too much into details, we state the following result that is extremely useful to the derivation of the analytical expressions of the two terms $\gamma_{k,2p}(\tau)$ and $\gamma_{k,2p-1}(\tau)$.\\
\indent \textbf{\textsc{Lemma 1}}: $u_k(\tau)$ and $v_k(\tau)$ are two \textit{independent} RVs whose distributions are given by:
\begin{eqnarray}
\label{pdf_u}\!\!\!\!p\big[u_k(\tau)\big]&=&\mathsmaller{\frac{{2\beta_{k,2p}}}{\sqrt{2\pi\sigma^2}}}~\!F_{k,2p}\big(u_k(\tau)\big)\mathlarger{e^{-\frac{u_k(\tau)^2}{2\sigma^2}}},\\
\label{pdf_v}\!\!\!\!p\big[v_k(\tau)\big]&=&\mathsmaller{\frac{{2\beta_{k,2p-1}}}{\sqrt{2\pi\sigma^2}}}~\!F_{k,2p-1}(v_k(\tau))\mathlarger{e^{-\frac{v_k(\tau)^2}{2\sigma^2}}}.
\end{eqnarray}
whith
\begin{eqnarray}\label{beta_2p}
\beta_{k,2p}&\triangleq\displaystyle&\displaystyle\prod_{l=1}^{p-1}\frac{1}{2\cosh\big(L_{2l}(k)/2\big)},\\
\label{beta_2p-1}\beta_{k,2p-1}&\triangleq\displaystyle&\displaystyle\prod_{l=1}^{p-1}\frac{1}{2\cosh\big(L_{2l-1}(k)/2\big)}.
\end{eqnarray}
\indent\textit{Proof:} see Appendix A.\\
As seen from (\ref{pdf_u}) and (\ref{pdf_v}), the two RVs  $u_k(\tau)$ and $v_k(\tau)$ are  \textit{almost} identically distributed (i.e., their pdfs have the same structure, but they are parametrized differently). Therefore, when evaluating the required expectation with respect to either $u_k(\tau)$ or $v_k(\tau)$, equivalent derivation steps can be followed to find either $\gamma_{k,2p}(\tau)$ or $\gamma_{k,2p-1}(\tau)$. As such,  we will only  derive  $\gamma_{k,2p}(\tau)$   and later deduce  the expression of $\gamma_{k,2p-1}(\tau)$ by easy identification. To that end, we denote the first and second derivatives of $F_{k,2p}(x)$  in (\ref{function_F}), with respect to the working variable $x$, by $F'_{k,2p}(x)$ and   $F''_{k,2p}(x)$, respectively.  We therefore establish the second partial derivative of $\ln\big(F_{k,2p}\big(u_k(\tau)\big)\big)$ with respect to the time delay parameter, $\tau$,  as follows:
\begin{eqnarray}
\mathsmaller{\frac{\partial^2}{\partial\tau^2}}\ln\big(\mathsmaller{F_{k,2p}}(\mathsmaller{\mathsmaller{u_{k}(\tau)}})\big)&=&
~\!\mathsmaller{\bm\dot{u}^2_{k}(\tau)}\!\left[\frac{\mathsmaller{F''_{k,2p}}(\mathsmaller{u_{k}(\tau)})}{\mathsmaller{F_{k,2p}}(\mathsmaller{u_{k}(\tau)})}-\frac{\mathsmaller{F'^{~\!\!2}_{k,2p}}(\mathsmaller{u_{k}(\tau)})}{\mathsmaller{F^{~\!\!2}_{k,2p}}(\mathsmaller{u_{k}(\tau)})}\!\right]+~\mathsmaller{\bm\ddot{u}_{k}(\tau)}\frac{\mathsmaller{F'_{k,2p}}(\mathsmaller{u_{k}(\tau)})}{\mathsmaller{F_{k,2p}}(\mathsmaller{u_{k}(\tau)})},\nonumber
\end{eqnarray} 
in which  $\bm\dot{u}_k(\tau)\triangleq\partial u_k(\tau)/\partial\tau$ and $\bm\ddot{u}_k(\tau)\triangleq\partial^2 u_k(\tau)/\partial\tau^2$.  We further show in Appendix A that $\bm\dot{u}_k(\tau)$ and $u_k(\tau)$ are two independent RVs as well. Thus, by applying the expectation operator to the previous equation, we obtain $\gamma_{k,2p}(\tau)$ as follows:
\begin{eqnarray}\label{second_derivative_of_F_expectation}
\!\!\!\!\gamma_{k,2p}(\tau)&\!\!\!=\!\!\!&\mathbb{E}\left\{\bm\dot{u}^2_{k}(\tau)\right\}\left[
\mathbb{E}\left\{\frac{\mathsmaller{F'^{~\!\!2}_{k,2p}}(\mathsmaller{u_{k}(\tau)})}{\mathsmaller{F^{~\!\!2}_{k,2p}}(\mathsmaller{u_{k}(\tau)})}\right\}-\mathbb{E}\left\{\frac{\mathsmaller{F''_{k,2p}}(\mathsmaller{u_{k}(\tau)})}{\mathsmaller{F_{k,2p}}(\mathsmaller{u_{k}(\tau)})}\!\right\}\right]~-~\mathbb{E}\left\{\!\bm\ddot{u}_{k}(\tau) \frac{\mathsmaller{F'_{k,2p}}(\mathsmaller{u_{k}(\tau)})}{\mathsmaller{F_{k,2p}}(\mathsmaller{u_{k}(\tau)})}\!\right\}.
\end{eqnarray}
In the sequel, we will derive  analytical the  expressions for the four expectations involved in (\ref{second_derivative_of_F_expectation}) separately. For convenience, we  define beforehand the following two quantities (for $q=2p$ and $2p-1$) that will appear repeatedly in the obtained expressions:
\begin{eqnarray}\label{omega}
\omega_{k,q} &\!\!\!\!\triangleq\!\!\!\!& 2~\!\beta_{k,q}\cosh\!\left(\!\mathsmaller{\frac{L_{q}(k)}{2}}\right)\sum_{i=1}^{2^{p-1}}\theta_{k,q}(i)d_{p}^{2}(2i\!-\!1)^{2},\\
\label{alpha}\alpha_{k,q}&\!\!\!\!\triangleq\!\!\!\!& 2~\!\beta_{k,q}\sinh\!\left(\!\mathsmaller{\frac{L_{q}(k)}{2}}\right)\sum_{i=1}^{2^{p-1}}\theta_{k,q}(i)d_{p}(2i-1).
\end{eqnarray}
\subsection{\textbf{Derivation of} $\mathbb{E}\big\{\bm\dot{u}_k^2(\tau)\big\}$}
\noindent In Appendix A, we show that :
\begin{eqnarray}\label{u_dot}
\!\!\!\!\!\!\!\!\!\!\!\!\bm{\dot}u_k(\tau)&\!\!\!\!=\!\!\!\!&\sqrt{E_s} \sum\displaylimits_{k'=1}^K \Re\big\{a(k')\big\}\bm \dot g\big([k-k']T\big)+\Re\big\{\bm{\dot}{w}_k(\tau)\big\},
\end{eqnarray}
where
\begin{equation}\label{derivative_noise}
\bm\dot w_k(\tau)=-\int_{-\infty}^{+\infty}\!\!\!w(t)\bm\dot h(t-kT-\tau)dt.
\end{equation}
Recall here that the transmitted symbols are assumed mutually independent. As they are also independent  from the derivative noise components and exploiting  the fact that $\mathbb{E}\big\{\bm\dot w_k(\tau)\big\}=0$ \big(since $\mathbb{E}\big\{ w(t)\big\}=0$\big), it can be shown that $\mathbb{E}\big\{\bm{\dot}u_k^2(\tau)\big\}$ is given by:
\begin{eqnarray}\label{u_dot_expectation}
\!\!\!\!\!\!\!\!\!\!\mathbb{E}\big\{\bm{\dot}u_k(\tau)^2\big\}&=&\mathsmaller{E_s}\left[\sum\displaylimits_{l=1}^K \mathbb{E}\Big\{\!\Re\big\{\!a(l)\!\big\}^2\!\Big\}\bm \dot g\big([k-l]T\big)^2\right.\nonumber\\
&&\left.+\sum\displaylimits_{l=1}^K\sum\displaylimits_{\mathclap{\substack{n=1\\n\neq l}} }^K\mathbb{E}\Big\{\Re\big\{a(l)\!\big\}\!\Big\}\mathbb{E}\Big\{\Re\big\{a(n)\!\big\}\!\Big\}\bm \dot g\big([k-l]T\big)\bm \dot g\big([k-n]T\big)\right]+~\mathbb{E}\Big\{\!\Re\big\{\!\bm{\dot}{w}_k(\tau)\!\big\}^2\!\Big\}.
\end{eqnarray}
The expected values of $\Re\big\{\!a(k)\!\big\}$ and  $\Re\big\{ a(k)\big\}^2$ involved in (\ref{u_dot_expectation}) are obtained by averaging them over all the points in the constellation alphabet, $\mathcal{C}_p$, i.e.:
\begin{eqnarray}\label{whole-alphabet}
\mathbb{E}\Big\{\Re\big\{a(k)\big\}^2\Big\}&\!\!\!\!\!=\!\!\!\!\!& \sum_{c_m\in\mathcal{C}_p}\!\!\! P\big[a(k)\!=\!{c}_m\big]\Re\big\{c_m\!\big\}^2.
\end{eqnarray}
\begin{eqnarray}\label{whole-alphabet_1}
\mathbb{E}\Big\{\Re\big\{a(k)\big\}\Big\}&\!\!\!\!\!=\!\!\!\!\!& \sum_{c_m\in\mathcal{C}_p}\!\!\! P\big[a(k)\!=\!{c}_m\big]\Re\big\{c_m\!\big\}.
\end{eqnarray}
Starting form (\ref{whole-alphabet}) and resorting to some algebraic manipulations, we show in Appendix B that:  
\begin{eqnarray}\label{simplified_real_square}
\!\!\!\!\!\!\!\!\!\!\mathbb{E}\Big\{\Re\big\{a(k)\big\}^2\Big\}&\!\!\!\!=\!\!\!\!&\omega_{k,2p}.
\end{eqnarray}
Using equivalent derivations, it can be also shown that:    
\begin{eqnarray}\label{simplified}
\!\!\!\!\!\!\!\!\!\!\mathbb{E}\Big\{\Re\big\{a(k)\big\}\Big\}&\!\!\!\!=\!\!\!\!&\alpha_{k,2p}.
\end{eqnarray}
%
In order to find the noise contribution through the derivative term in  (\ref{u_dot_expectation}), we recall that the original continuous-time noise is assumed to be white, i.e.,  $\mathbb{E}\big\{\!\Re\big\{\!{w}(t_1)\!\big\}\Re\big\{\!{w}(t_2)\!\big\}\!\big\}=\sigma^2\delta(t_1-t_2)$. Therefore, starting from the expression of $\bm{\dot}{w}_k(\tau)$ in (\ref{derivative_noise}) and resorting to equivalent manipulations as in (\ref{E_U_dU}) of Appendix A, we obtain:
\begin{eqnarray}\label{noise_conntribution}
\mathbb{E}\Big\{\Re\big\{\bm{\dot}{w}_k(\tau)\big\}^2\Big\}&\!\!\!\!=\!\!\!\!&\sigma^2\int_{\mathbb{R}}\bm\dot h\big(t-kT-\tau\big)^2~\!dt,\nonumber\\
&\!\!\!\!=\!\!\!\!&-\sigma^2\int_{\mathbb{R}} h\big(t-kT-\tau\big)\bm\ddot  h\big(t-kT-\tau\big)dt,\nonumber\\
&\!\!\!\!=\!\!\!\!&-\sigma^2 \bm\ddot g(0).
\end{eqnarray}
Note here that, in line with the left-hand side of (\ref{noise_conntribution}), the right-hand side of the same equation is indeed positive  since $\bm\ddot g(0)<0$. This is because the filter $g(.)$ is convex in the vicinity of zero where it also attains its maximum. Now, using (\ref{simplified_real_square}) to (\ref{noise_conntribution}) in (\ref{u_dot_expectation}), it can be easily shown that:
\begin{eqnarray}\label{second_expectation}
\!\!\!\!\!\!\!\!\!\mathbb{E}\big\{\bm\dot{u}_k(\tau)^2\big\}&\!\!\!\!=\!\!\!\!&\mathsmaller{E_s}\sum\displaylimits_{l=1}^K\Big(\omega_{l,2p}-\alpha_{l,2p}^2\Big)\bm \dot g^2\big([l-k]T\big)+\mathsmaller{E_s}\!\left(\sum_{l=1}^K\alpha_{l,2p}~\!\bm \dot g\big([l\!-\!k]T\big)\!\!\right)^{\!2}\!\!-\sigma^2 \bm\ddot g(0).
\end{eqnarray}

\subsection{\textbf{Derivation of} $\mathbb{E}\left\{
\left(F'_{k,2p}\big(u_k(\tau)\big)\big/F_{k,2p}\big(u_k(\tau)\big)\right)^2\right\}$}
This is nothing but  the expected value of a known transformation of the RV, $u_k(\tau)$, whose distribution was already established in (\ref{pdf_u}). Therefore, it can be evaluated in closed form by  integration over $p\big[u_k(\tau)\big]$ as follows: 
\begin{eqnarray}\label{integral_0}
\mathbb{E}\mathsmaller{\left\{\left(\frac{F'_{k,2p}\big(u_k(\tau)\big)}{F_{k,2p}\big(u_k(\tau)\big)}\right)^2\right\}}&\!\!\!\!\!=\!\!\!\!\!&\int_{\mathbb{R}} \mathsmaller{ \frac{F'^2_{k,2p}\big(u_k(\tau)\big)}{F^2_{k,2p}\big(u_k(\tau)\big)}}p[u_k(\tau)] \mathrm du_k(\tau)\nonumber\nonumber\\
&\!\!\!\!\!=\!\!\!\!\!&\mathsmaller{\frac{{2\beta_{k,2p}}}{\sqrt{2\pi\sigma^2}}}\!\int_{\mathbb{R}} \! \mathsmaller{ \frac{F'^2_{k,2p}\big(u_k(\tau)\big)}{F_{k,2p}\big(u_k(\tau)\big)}}\mathlarger{e^{\frac{-u_k(\tau)^2}{2\sigma^2}}} \mathrm du_k(\tau).\nonumber
\end{eqnarray}
After using the explicit expression of $F'_{k,2p}\big(u_k(\tau)\big)$, the last equality is  further simplified by using the variable substitution $t=\sqrt{2}u_k(\tau)/\sigma$ to obtain:
\begin{eqnarray}\label{fourth_expectation}
\mathbb{E}\mathsmaller{\left\{\left(\frac{F'_{k,2p}\big(u_k(\tau)\big)}{F_{k,2p}\big(u_k(\tau)\big)}\right)^2\right\}}&=&\frac{2\rho}{\sigma^2}~\!\Psi_{k,2p}(\rho),
\end{eqnarray}
where $\Psi_{k,2p}(.)$ in the last equality is given  by:
\begin{eqnarray}\label{integral_psi}
\Psi_{k,2p}(\rho)&=&\mathsmaller{\frac{\beta_{k,2p}~\!d_{p}^{2}}{\sqrt{\mathlarger{\pi}}}}\int_{-\infty}^{+\infty}\frac{\lambda_{k,2p}^2(t,\rho)}{\delta_{k,2p}(t,\rho)}\hspace{0.2cm}\mathlarger{e^{-\frac{t^2}{4}}}dt,
\end{eqnarray}
with
\begin{eqnarray}
\lambda_{k,2p}(t,\rho)&=&\displaystyle\sum_{i=1}^{2^{p-1}}(2i-1)\theta_{k,2p}(i)e^{\!-[2i-1]^2d_p^2\rho}\times\mathsmaller{\sinh\bigg(\!\!\sqrt{ \rho}[2i\!-\!1]d_pt+\frac{L_{2p}(k)}{2}\!\bigg)},\nonumber
\end{eqnarray}
\begin{eqnarray}
\delta_{k,2p}(t,\rho)&=&\!\displaystyle\sum_{i=1}^{2^{p-1}}\!\theta_{k,2p}(i)e^{-[2i-1]^2d^2_p\rho}\mathsmaller{\cosh\!\bigg(\!\!\sqrt{\rho}[2i\!-\!1]d_pt+\frac{L_{2p}(k)}{2}\!\bigg)}.\nonumber
\end{eqnarray}
\subsection{\textbf{Derivation of} $~\mathbb{E}\left\{
F''_{k,2p}\big(u_k(\tau)\big)\big{/}F_{k,2p}\big(u_k(\tau)\big)\right\}$}
This expectation can also be explicitly found by integrating over  $p\big[u_k(\tau)\big]$ in (\ref{pdf_u}) to yield:
\begin{eqnarray}\label{integral}
\!\!\!\!\!\!\!\!\!\!\mathbb{E}\left\{\mathsmaller{\frac{F''_{k,2p}\big(u(k)\big)}{F_{k,2p}\big(u_k(\tau)\big)}}\right\}&=&\int_{\mathbb{R}}\mathsmaller{ \frac{F''_{k,2p}\big(u_k(\tau)\big)}{F_{k,2p}\big(u_k(\tau)\big)}}p[u_k(\tau)] \mathrm du_k(\tau)\nonumber\\
\!\!\!\!\!\!\!\!\!\!&\!\!\!\!\!=\!\!\!\!\!&\mathsmaller{\frac{{2\beta_{k,2p}}}{\sqrt{2\pi\sigma^2}}}\!\int_{\mathbb{R}} \! \textstyle{F''_{k,2p}\big(u_k(\tau)\big)}\mathlarger{e^{\frac{-u_k(\tau)^2}{2\sigma^2}}} \mathrm du_k(\tau),
\end{eqnarray}
in which the second derivative of the function $F_{k,2p}(.)$ defined in (\ref{function_F}) is given by:
\begin{eqnarray}\label{second_derivative_of_F_explicit}
\!\!\!\!\!\!\!\!F''_{k,2p}(x)&=&\mathsmaller{\frac{E_sd_p^2}{\sigma^4}}\sum_{i=1}^{2^{p-1}}\textstyle (2i-1)^2 \theta_{k,2p}(i)\mathlarger{e^{-\rho[2i-1]^2d_p^2}}\times\mathsmaller{\cosh\bigg(\!\frac{\sqrt{E_s}[2i-1]d_p}{\sigma^2}x+\frac{L_{2p}(k)}{2}\!\bigg)}.
\end{eqnarray}
After expanding (\ref{second_derivative_of_F_explicit}) using the identity $\cosh(x+y)=\cosh(x)\cosh(y)+\sinh(x)\sinh(y)$, plugging the result back into (\ref{integral}) and then using the fact that $\sinh(x)e^{-\frac{x^2}{2}}$ is an odd function (i.e., its integral is identically zero), it follows that (\ref{integral}) is explicitly given by:
\begin{eqnarray}\label{integral_1}
\mathbb{E}\left\{\mathsmaller{\frac{F''_{k,2p}\big(u(k)\big)}{F_{k,2p}\big(u_k(\tau)\big)}}\right\}&\!\!\!\!\!\!\!\!\!\!&\nonumber\\
&\!\!\!\!\!\!\!\!\!\!&\!\!\!\!\!\!\!\!\!\!\!\!\!\!\!\!\!\!\!\!\!\!\!\!\!\!\!\!\!\!\!\!\!\!\!\!\!\!\!\!=\mathsmaller{\frac{{2\beta_{k,2p}}E_sd_p^2}{\sqrt{2\pi\sigma^2}\sigma^4}}\mathsmaller{\cosh\left(\frac{L_{2p}(k)}{2}\right)}\sum_{i=1}^{2^{p-1}} \mathsmaller{(2i-1)^2 \theta_{k,2p}^{(i)}\mathlarger{e^{-\rho[2i-1]^2d_p^2}}}\int_{-\infty}^{+\infty} \!\! \mathsmaller{\cosh\bigg(\!\frac{\sqrt{E_s}[2i-1]d_p}{\sigma^2}u_k(\tau)\!\bigg)}\mathlarger{e^{\frac{-u_k(\tau)^2}{2\sigma^2}}} \mathrm du_k(\tau).\nonumber\\
&& 
\end{eqnarray}
Moreover, we show  via ``integration by parts'', the following equality for any $a>0$ and $b \in \mathbb{R}$:
\begin{eqnarray}
\label{formula_cosh}\int_{0}^{+\infty}\!\!\!\cosh\big(b~\!x\big)\mathlarger{e^{-ax^2}}dx&\!\!=\!\!&\mathsmaller{\frac{1}{2}\sqrt{\frac{\pi}{a}}}\mathlarger{e^{\frac{b^2}{4a}}},
\end{eqnarray}
which  is used in (\ref{integral_1}), with the appropriate identifications, to yield the following closed-form expression for the expectation in (\ref{integral}):
\begin{eqnarray}\label{integral_of_second_derivative_closed}
\mathbb{E}\left\{\mathsmaller{\frac{F''_{k,2p}\big(u_k(\tau)\big)}{F_{k,2p}\big(u_k(\tau)\big)}}\right\}&\!\!\!\!=\!\!\!\!&\frac{2\omega_{k,2p}}{\sigma^2}\rho.
\end{eqnarray} 

\subsection{\textbf{Derivation of} $~\mathbb{E}\left\{\bm\ddot u_k(\tau)F'_{k,2p}\big(u_k(\tau)\big)\big{/}F_{k,2p}\big(u_k(\tau)\big)\right\}$}
To find this expectation, we use a standard  approach in which we first find  the expectation   conditioned on $u_k(\tau)$ and then  average the obtained result  with respect to $u_k(\tau)$. By doing so, we obtain:
\begin{eqnarray}\label{u_snd_f_prime_f_expectation}
\!\!\!\!\!\!\!\!\!\!\!\!\mathbb{E}\!\left\{\!\bm\ddot{u}_k(\tau) \mathsmaller{\frac{F'_{k,2p}\big(u_k(\tau)\big)}{F_{k,2p}\big(u_k(\tau)\big)}}\right\}
&=&\mathbb{E}_{u_k}\!\Bigg\{\!\mathsmaller{\mathbb{E}\Big\{\!\bm\ddot{u}_k(\tau) \big|u_k(\tau)\!\Big\}\frac{F'_{k,2p}\big(u_k(\tau)\big)}{F_{k,2p}\big(u_k(\tau)\big)}} \!\Bigg\}\!.
\end{eqnarray}
In order to find $\mathbb{E}_{\bm\ddot{u}_k|u_k}\!\big\{\!\bm\ddot{u}_k(\tau) \big|u_k(\tau)\!\big\}$ in (\ref{u_snd_f_prime_f_expectation}), we must find the explicit expression of  $\bm\ddot u_k(\tau)$ as function of $u_k(\tau)$. In fact, it is easy to show that:
\begin{eqnarray}\label{u_k_second_derivative}
\!\!\!\!\bm\ddot{u}_k(\tau)&\!\!\!=\!\!\!&\int_{-\infty}^{+\infty}\Re\big\{y(t)\big\}\bm\ddot{h}(t-kT-\tau)dt\nonumber\\
\!\!\!\!&\!\!\!=\!\!\!&\mathsmaller{\sqrt{E_s}}\sum_{l=1}^{K}\Re\big\{a(l)\big\}\bm\ddot{g}\big([l-k]T\big)+\Re\big\{\bm\ddot{w}_k(\tau)\big\}.
\end{eqnarray}
Moreover, from (\ref{y_filterd}) and (\ref{z_k_simplified}) in Appendix A, we readily have:
\begin{eqnarray}\label{u_k_alone}
u_l(\tau)=\sqrt{E_s}~\!\Re\{a(l)\}+\Re\{ {w}_l(\tau)\}.
\end{eqnarray}
Therefore,  $\Re\{a(l)\}=\frac{1}{\sqrt{E_s}}\left[u_l(\tau)-\Re\{ {w}_l(\tau)\}\right]$ 
which is used in (\ref{u_k_second_derivative}) to obtain:
\begin{eqnarray}\label{u_k_second_derivative_simplif}
\!\!\!\!\!\!\!\!\bm\ddot{u}_k(\tau)&=&\!\sum\displaylimits_{l=1}^K  \Big[u_{l}(\tau)\!-\!\Re\big\{{w}_{l}(\tau)\big\}\Big]\bm\ddot{g}\big([l\!-\!k]T\big)+\Re\big\{\bm\ddot{w}_k(\tau)\big\}.
\end{eqnarray}
Now, since $\mathbb{E}\big\{\Re\{\bm\ddot{w}_k(\tau)\}\big\}=\mathbb{E}\big\{\Re\{w_k(\tau)\}\big\}=0$ and since the RVs $\{u_l(\tau)\}_l$ are mutually independent, it follows that:
\begin{eqnarray}\label{expectaion_u_second_given_u}
\!\!\!\!\!\!\!\!\!\mathbb{E}\big\{\bm\ddot{u}_k(\tau) \big|u_k(\tau)\big\}&=&{u}_k(\tau)\bm\ddot{g}(0)+\sum\displaylimits_{\mathclap{\substack{l=1\\l\neq k}} }^K  \mathbb{E}\big\{u_{l}(\tau)\big\}~\!\bm\ddot{g}\big([l\!-\!k]T\big).
\end{eqnarray}
But owing to (\ref{u_k_alone}) and (\ref{simplified}), it immediately follows that:
\begin{eqnarray}\label{expectaion_u_l_over_u_second}
\mathbb{E}\big\{u_{l}(\tau)\big\}=\mathsmaller{\sqrt{E_s}}~\!\mathbb{E}\big\{\Re\{a(l)\}\big\}=\mathsmaller{\sqrt{E_s}}~\!\alpha_{k,2p}.
\end{eqnarray}
Using (\ref{expectaion_u_l_over_u_second}) in (\ref{expectaion_u_second_given_u}) and then plugging the obtained result back into (\ref{u_snd_f_prime_f_expectation}), we obtain:
\begin{eqnarray}\label{u_snd_f_prime_f_expectation_final}
\!\!\!\!\!\!\!\!\!\!\mathsmaller{\mathbb{E}\left\{\bm\ddot{u}_k(\tau) \frac{F'_{k,2p}\big(u_k(\tau)\big)}{F_{k,2p}\big(u_k(\tau)\big)}\right\}}&=&\bm\ddot{g}\big(0\big)\mathsmaller{\mathbb{E}\left\{{u}_k(\tau)\frac{F'_{k,2p}\big(u_k(\tau)\big)}{F_{k,2p}\big(u_k(\tau)\big)}\right\}}
+\mathsmaller{\sqrt{E_s}}~\!\mathsmaller{\mathbb{E}\left\{\frac{F'_{k,2p}\big(u_k(\tau)\big)}{F_{k,2p}\big(u_k(\tau)\big)}\right\}}\!\sum\displaylimits_{\mathclap{\substack{l=1\\l\neq k}} }^K \alpha_{l,2p}~\!\bm\ddot{g}\big([l-k]T\big).
\end{eqnarray}
As done previously, the two expectations in (\ref{u_snd_f_prime_f_expectation_final}) are derived in closed-form by integration over the distribution, $p\big[u_k(\tau)\big]$, already established in (\ref{pdf_u}). The final results are given by:
\begin{eqnarray}\label{u_f_prime_f_expectation}
\mathsmaller{\mathbb{E}\left\{{u}_k(\tau)\frac{F'_{k,2p}\big(u_k(\tau)\big)}{F_{k,2p}\big(u_k(\tau)\big)}\right\}}&\!\!\!=\!\!\!&\mathsmaller{2\rho~\!\omega_{k,2p}},\\
\label{integral_of_first_derivative_closed}
\mathbb{E}\left\{\mathsmaller{\frac{F'_{2p,\bm{\alpha}}\big(u(k)\big)}{F_{2p,\bm{\alpha}}\big(u_k(\tau)\big)}}\right\}&\!\!\!=\!\!\!&\mathsmaller{\frac{\sqrt{E_s}}{\sigma^2}~\!\alpha_{k,2p}},
\end{eqnarray}
which are used in  (\ref{u_snd_f_prime_f_expectation_final}) to yield:
\begin{eqnarray}\label{u_snd_f_prime_f_expectation_final_final}
\!\!\!\!\!\!\!\!\!\!\mathsmaller{\mathbb{E}\left\{\bm\ddot{u}_k(\tau)\frac{F'_{k,2p}\big(u_k(\tau)\big)}{F_{k,2p}\big(u_k(\tau)\big)}\right\}}&=&~\!2\rho\bigg[\omega_{k,2p}~\!\bm\ddot{g}\big(0\big)
+\!\alpha_{k,2p}\displaystyle\sum_{l\neq k}\alpha_{l,2p}~\!\bm\ddot{g}\big([l-k]T\big)\bigg].
\end{eqnarray}
Finally, by injecting (\ref{second_expectation}), (\ref{fourth_expectation}), (\ref{integral_of_second_derivative_closed}),  and (\ref{u_snd_f_prime_f_expectation_final_final}) back into (\ref{second_derivative_of_F_expectation}), the analytical expression of $\gamma_{k,2p}(\tau)$ is obtained as:
\begin{eqnarray}\label{gamma_k_2p_final}
\!\!\!\!\!\!\!\!\gamma_{k,2p}(\tau)&=&4\rho^2\Big[\omega_{k,2p}-\Psi_{k,2p}(\rho)\Big]
\left[\sum\displaylimits_{l=1}^K\Big(\omega_{l,2p}-\alpha_{l,2p}^2\Big)\bm \dot g^2\big([l-k]T\big)+\left(\sum_{l=1}^K\alpha_{l,2p}~\!\bm \dot g\big([l\!-\!k]T\big)\!\!\right)^{\!2}\right]\nonumber\\
&\!\!\!\!\!\!\!\!&~~~~~~~~~~~~~~~~~~~~~~~~~~~~~~~~~~~~~~~~~-2\rho\bigg[\Psi_{k,2p}(\rho)~\!\bm\ddot{g}\big(0\big)-\alpha_{k,2p}\displaystyle\sum_{l\neq k}\alpha_{l,2p}~\!\bm\ddot{g}\big([l-k]T\big)\bigg].
\end{eqnarray}
Due to the apparent symmetries between the distributions of the two RVs $u_k(\tau)$ and $v_k(\tau)$, the analytical expression of $\gamma_{k,2p-1}(\tau)$ can be directly deduced from the one of $\gamma_{k,2p}(\tau)$ by easy identifications as:
\begin{eqnarray}\label{gamma_k_2p_final}
\label{gamma_k_2p-1_final}\gamma_{k,2p-1}(\tau)&\!\!\!\!=\!\!\!\!&4\rho^2\Big[\omega_{k,2p-1}-\Psi_{k,2p-1}(\rho)\Big]
\left[\sum\displaylimits_{l=1}^K\Big(\omega_{l,2p-1}-\alpha_{l,2p-1}^2\Big)\bm \dot g^2\big([l-k]T\big)+\left(\sum_{l=1}^K\alpha_{l,2p-1}~\!\bm \dot g\big([l\!-\!k]T\big)\!\!\right)^{\!2}\right]\nonumber\\
&\!\!\!\!\!\!\!\!&~~~~~~~~~~~~~~~~~~~~~~~~~~~~~~~~~~~~~~-2\rho\bigg[\Psi_{k,2p-1}(\rho)~\!\bm\ddot{g}\big(0\big)+\alpha_{k,2p-1}\displaystyle\sum_{l\neq k}\alpha_{l,2p-1}~\!\bm\ddot{g}\big([l-k]T\big)\bigg].
\end{eqnarray}
The closed-form expression for the TD CA CRLB is then obtained as the inverse of the Fisher information given by (\ref{Fisher_dinformation_1}), i.e.:
\begin{eqnarray}\label{CRLB_final}
\textrm{CRLB}(\tau)&\!\!\!=\!\!\!&\frac{1}{\textstyle\sum_{k=1}^{K}\gamma_{k,2p}(\tau)+\gamma_{k,2p-1}(\tau)}.
\end{eqnarray}
It is worth mentioning  here that the turbo-code setup  is not needed in our derivations and that the new CA CRLB expression (\ref{CRLB_final}) is actually valid for any coded system in general. In fact, we have so far  only  exploited the fact that the constellation is Gray-coded   and we have expressed the CA TD CRLBs  explicitly in terms of the coded bits' \textit{a priori} LLRs. Yet, we will explain in the next subsection how these unknown LLRs are obtained from the output of the SISO decoders in a turbo-coded system. Yet, they can also be obtained from LDPC-coded systems in the very same way if the latter are decoded with the turbo principle [\ref{LDPC_2}], [\ref{LDPC_1}]  (i.e., MAP or BCJR decoder). In this case, the so-called check nodes (C-nodes) and variable nodes (V-nodes) [\ref{LDPC_2}] play the very same role as SISO decoders in turbo-coded systems.
%

 \subsection{Evaluation of the analytical  CA CRLBs}\label{CRLB_evaluation}
In order to compute and plot the new CA CRLBs, one needs to evaluate the coefficients $\omega_{k,q}$ and $\alpha_{k,q}$ for $q=2p$ and $q=2p-1$. These coefficients are, however, functions of the \textit{a priori} LLRs, $L_l(k)$, as seen from (\ref{omega}) and (\ref{alpha}). In the sequel, we briefly explain how these LLRs can be obtained from the output of the SISO decoders at the convergence of the BCJR algorithm.
First, the MF returns a  sequence of $K$ symbol-rate samples:
\begin{equation}\label{seq_CRLB}
\mathbf{y}(\tau)~\!=~\!\big[y_1(\tau),y_2(\tau),\ldots, y_{K}(\tau)\big]^T,
\end{equation}
where (cf. Appendix A):
\begin{eqnarray}\label{symbol-rate}
\!\!\!\!\!\!\!\!y_k(\tau)&\!\!\!\!=\!\!\!\!&\!\!\int_{-\infty}^{+\infty}\!\!\!y(t)h(t-kT-\tau)dt~\!=~\!\sqrt{E_s}~\!a(k)+ w_k(\tau).
\end{eqnarray}
Then, the soft demapper extracts the so-called \textit{bit likelihoods}:
\begin{eqnarray}\label{bit_likelihoods_CRLB}
\Lambda_l(k)&\!\!\triangleq\!\!&\ln\left(\frac{p\big[\mathbf{y}(\tau)\big|b_l^k=1\big]}
{p\big[\mathbf{y}(\tau)\big|b_l^k=0\big]}\right),
\end{eqnarray}
for all the code bits and feed them as inputs to the turbo decoder. By exchanging the so-called \textit{extrinsic} information between the two SISO decoders, the \textit{a posteriori} LLRs of the code bits:
\begin{eqnarray}\label{a_posteriori_LLRs}
\Upsilon_{l}(k)=\ln\left(\frac{P\big[b_l^k=1\big|\mathbf{y}(\tau)\big]}{P\big[b_l^k=0\big|\mathbf{y}(\tau)\big]}\right).
\end{eqnarray}
are updated iteratively according to the turbo principle. We denote their values at the $r^{th}$ turbo iteration as $\Upsilon_l^{(r)}(k)$.
After say $R$ turbo iterations, a steady state is achieved wherein $\Upsilon_l^{(R)}(k)\approx\Upsilon_l(k)$, for every $l$ and $k$, and their signs are used to detect the bits. Yet, owing to the well-known Bayes' formula, we have:
\begin{eqnarray}\label{Bayes_1}
P\big[b_l^k=1\big|\mathbf{y}(\tau)\big]=\frac{p\big[\mathbf{y}(\tau)\big|b_l^k=1\big]P\big[b_l^k=1\big]}{p[\mathbf{y}(\tau)]},
\end{eqnarray}
and
\begin{eqnarray}\label{Bayes_0}
P\big[b_l^k=0\big|\mathbf{y}(\tau)\big]=\frac{p\big[\mathbf{y}(\tau)\big|b_l^k=0\big]P\big[b_l^k=0\big]}{p[\mathbf{y}(\tau)]}.
\end{eqnarray}
Therefore, by taking the ratio of (\ref{Bayes_1}) and (\ref{Bayes_0}) and applying the natural logarithm, it immediately follows that:
\begin{eqnarray}\label{apriori-aposteriori}
L_l(k)=\Upsilon_l(k)-\Lambda_l(k)\approx\Upsilon_l^{(R)}(k)-\Lambda_l(k),
\end{eqnarray}
meaning that the required \textit{a priori} LLRs of the code bits can be easily obtained from their steady-state \textit{a posteriori} LLRs and $\Lambda_l(k)$ already computed by the \textit{soft} demapper prior to data decoding. 
\section{New Time Delay CA ML estimator}
\begin{figure}
\hskip 4.5cm
\vskip -4cm
\begin{centering}
\includegraphics[width=\textwidth]{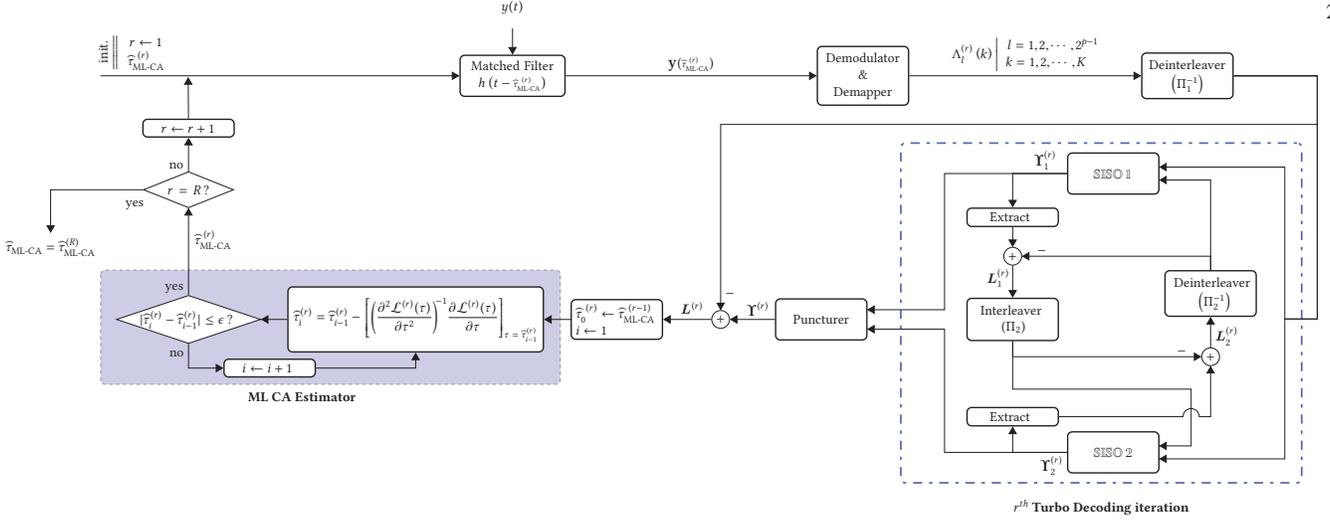}
\end{centering}
\vspace{-5cm}
\caption{Flowchart of the new  CA TD ML estimator.}
\label{flow_chart}
\end{figure}

As mentioned previously, the timing recovery  task is integrated within the turbo iteration loop. 
But in order to initiate the turbo decoding process itself, the latter  needs some preliminary information-bearing symbol-rate samples. The latter can be obtained at the output of the MF  \big(corrected with $\widehat{\tau}_\textrm{ML-NDA}$\big) where $\widehat{\tau}_\textrm{ML-NDA}$ is the NDA MLE for the TD parameter estimated as: 
\begin{eqnarray}\label{E24}
\widehat{\tau}_{\textrm{ML-NDA}}\!~=~\!\underset{\tau}\argmax\!~\mathcal{L}^{(0)}(\tau),
\end{eqnarray}
where $\mathcal{L}^{(0)}(.)$ is the NDA LLF obtained directly from its CA counterpart in (\ref{Final_LLF}) by setting\footnote{In the NDA case
(i.e., before starting data decoding), no \textit{a priori} information
about the bits is available at the receiver end, i.e., $P[b_l^k=0]=P[b_l^k=1]=1/2$ and thus $L_{l}(k)=0$ for all $l$ and $k$.} $L_{l}(k)=0$  for all $l$ and $k$, i.e.:
\begin{eqnarray}\label{E25}
\mathcal{L}^{(0)}(\tau)=\sum_{k=0}^{K-1}\!\Big[\ln\big(\mathsmaller{F\big(u_{k}(\tau)\big)}\big)\!+\!\ln\big(\mathsmaller{F\big(v_{k}(\tau)\big)}\big)\Big],
\end{eqnarray}
in which $F(.)$ is simply given by:
\begin{eqnarray}
\!\!F(x)&\!\!\!\!=\!\!\!\!&\!\!\sum_{i=1}^{2^{p-1}}\textstyle \!\mathlarger{e^{-\rho N_ad_p^2[2i-1]^2}}\cosh\Big(\frac{2S[2i-1]\sqrt{N_a}d_p}{\sigma^2}\!~x\Big).\nonumber
\end{eqnarray}
The iterative algorithm that maximizes $\mathcal{L}^{(0)}(\tau)$ with respect to $\tau$ in (\ref{E24}) will be detailed at the end of this section. Note also  that $u_k(\tau)$ and $v_k(\tau)$  involved in  (\ref{E25}) are the real and imaginary parts of a discrete-time MF output that is obtained as follows.
At the receiver side, $y(t)$ is upsampled using a sampling period $T_s<T/(1+\beta)$ with $\beta$ being the roll-off factor to obtain:
\begin{equation}\label{output_samples}
y_l\triangleq y(lT_s)= \sqrt{E_s}~\sum_{k=1}^{K}a(k)~h(lT_s-kT-\tau)+w(lT_s).\nonumber
\end{equation}
These high-rate samples are then passed through a discrete-time MF to obtain the  symbol-rate samples:
\begin{eqnarray}\label{u_and_v_using_sums}
y_k(\tau)&\!\!\!\!=\!\!\!\!&y_l\star h(lT_s\!-\!kT\!-\!\tau)=\sum_{l}~\! y_l\!~h(lT_s\!-\!kT\!-\!\tau)dt,\nonumber
\end{eqnarray}
from which we obtain $u_k(\tau)=\Re\{y_k(\tau)\}$ and $v_k(\tau)=\Im\{y_k(\tau)\}$ which are used in (\ref{E25}). Once $\widehat{\tau}_{\textrm{ML-NDA}}$ is acquired, the corresponding sequence of symbol-rate samples:
\begin{eqnarray}\label{symbol-rate_NDA}
\mathbf{y}(\widehat{\tau}_{\textrm{ML-NDA}})~\!=~\!\big[y_1(\widehat{\tau}_{\textrm{ML-NDA}}),y_2(\widehat{\tau}_{\textrm{ML-NDA}}),\ldots, y_{K}(\widehat{\tau}_{\textrm{ML-NDA}})\big]^T\!\!,\nonumber
\end{eqnarray} 
is passed to the soft demapper in order to find the \textit{bit likelihoods} required to start the decoding process. To exploit the output of the decoder and better re-synchronize the system, at a \textit{per-turbo-iteration} basis, we modify (\ref{apriori-aposteriori}) as follows:
\begin{eqnarray}\label{apriori-aposteriori_r}
L_l^{(r)}(k)=\Upsilon_l^{(r)}(k)-\Lambda_l^{(r-1)}(k),
\end{eqnarray}
in order to obtain a more refined TD estimate, $\widehat{\tau}^{(r)}_\textrm{ML-CA}$, after each $r^{th}$ turbo iteration as will be explained shortly. Note here that  $\Lambda_l^{(r-1)}(k)$ are the bit likelihoods that are obtained after re-synchronizing the system with $\widehat{\tau}^{(r-1)}_\textrm{ML-CA}$, i.e., the TD estimate corresponding to the previous turbo iteration. These are fed to the SISO decoders to compute an update for the \textit{a posteriori} LLRs, $\Upsilon_l^{(r)}(k)$, at the current $r^{th}$ turbo iteration. 
The refined TD MLE is thereof obtained as:
\begin{eqnarray}\label{TD_CA_MLE_r}
\widehat{\tau}_{\textrm{ML-CA}}^{(r)}&=&\underset{\tau}{\argmax}~\mathcal{L}^{(r)}(\tau),
\end{eqnarray}
where $\mathcal{L}^{(r)}(\tau)$ is the CA LLF in (\ref{Final_LLF}) evaluated using $L_l^{(r)}(k)$  instead of $L_l(k)$, i.e.: 
\begin{eqnarray}\label{LLF_r}
\!\!\!\!\!\!\!\!\!\!\!\mathcal{L}^{(r)}(\tau)&=&\sum_{k=1}^{K}\ln\Big(F^{(r)}_{k,2p}\big(u_k(\tau)\big)\Big)+\ln\Big(F^{(r)}_{k,2p-1}\big(v_k(\tau)\big)\Big),\nonumber
\end{eqnarray}
in which $F^{(r)}_{k,q}(.)$ is given by:
\begin{eqnarray}\label{function_F_r}
F^{(r)}_{k,q}(x)&=&\!\sum_{i=1}^{2^{p-1}}\textstyle\! \theta_{k,q}^{(r)}(i)\mathlarger{e^{-\rho d_p^2[2i-1]^2}}\cosh~\!\!\!\bigg(\!\frac{\sqrt{E_s}[2i-1]d_p}{\sigma^2}x+\frac{L^{(r)}_{q}(k)}{2}\!\bigg),\nonumber
\end{eqnarray}
for $q=2p$ and $2p\!-\!1$. Here, $\widehat{\theta}_{k,2p}^{(r)}(i)$ and $\widehat{\theta}_{k,2p-1}^{(r)}(i)$ are also obtained by using $L_l^{(r)}(k)$ instead of $L_l(k)$ in (\ref{theta_2p}) and (\ref{theta_2p-1}), respectively.
\\A key detail  that is still missing  needs to be addressed here as how the  NDA and CA LLFs  are  maximized in (\ref{E24}) and (\ref{TD_CA_MLE_r}). Actually, since these LLFs were derived in closed-form expressions, they can be easily maximized using any of the popular iterative techniques such as the well-known Newton-Raphson algorithm:
\begin{eqnarray}\label{Neton_Raphson_formula}
\!\!\!\!\!\!\!\!\!\!\!\!\widehat{\tau}_{i}^{(r)}&=&\widehat{\tau}_{i-1}^{(r)}-\left[{\left(\frac{\partial^2\mathcal{L}^{(r)}(\tau)}{\partial \tau^2}\right)}^{\!\!-1} \frac{\partial\mathcal{L}^{(r)}(\tau) }{\partial \tau} \right]_{\tau~\!=~\!\widehat{\tau}_{i-1}^{(r)}},
\end{eqnarray}
in which $\widehat{\tau}_{i}^{(r)}$ is the TD update pertaining to the $i^{th}$ Newton-Raphson iteration. The algorithm stops once the convergence criterion $|\mathsmaller{\widehat{\tau}_{i}^{(r)}-\widehat{\tau}_{i-1}^{(r)}}|\leq\epsilon$ is met\footnote{Note here that $\epsilon$ is a predefined threshold that governs the required estimation accuracy.} to produce $\widehat{\tau}_{\textrm{ML-CA}}^{(r)}$ as the CA TD MLE during the $r^{th}$ turbo iteration. Note, however, that the Newton-Raphson algorithm itself is iterative in nature and, therefore, requires  a reliable initial guess, $\widehat{\tau}_{0}^{(r)}$, to ensure its convergence to the global maximum of the underlying objective LLF.  At each $r^{th}$ turbo iteration,  the algorithm is initialized by $\widehat{\tau}_{0}^{(r)}=\widehat{\tau}_{\textrm{ML-CA}}^{(r-1)}$ (i.e., by the TD MLE pertaining to the previous turbo iteration). At the very first turbo iteration,  however, the algorithm is initialized with the NDA MLE, $\widehat{\tau}_{\textrm{ML-NDA}}$, obtained in (\ref{E24}). The latter is obtained by maximizing $\mathcal{L}^{(0)}(\tau)$  itself via the very same Newton-Raphson algorithm and the corresponding initial guess is obtained by a broad line search over $\tau$. For better illustration, Fig. \ref{flow_chart} depicts the architecture of the newly proposed CA ML timing recovery algorithm. \\
\section{Simulation Results}\label{section_5}
\begin{figure}[!h] 
 \vskip 0.25 cm 
\begin{center}
\includegraphics[scale = 0.35] {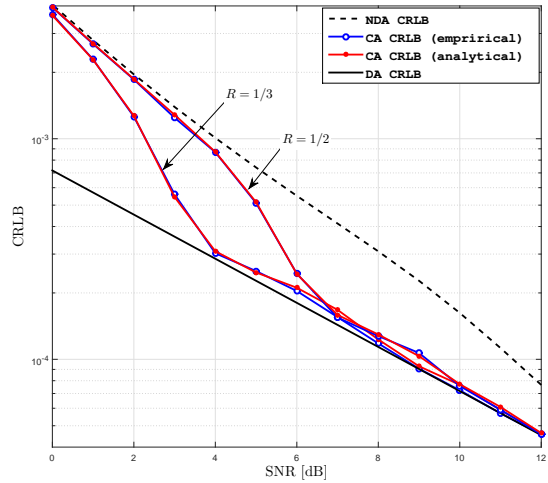}
\end{center}
\caption{Comparison between the empirical and analytical CA CRLBs for different code rates, $R$, as function of the SNR: 16-QAM, rolloff $=0.2$.}
\label{crlb-16-QAM}
\end{figure}

\begin{figure}[!ht] 
\begin{center} 
\includegraphics[scale = 0.36] {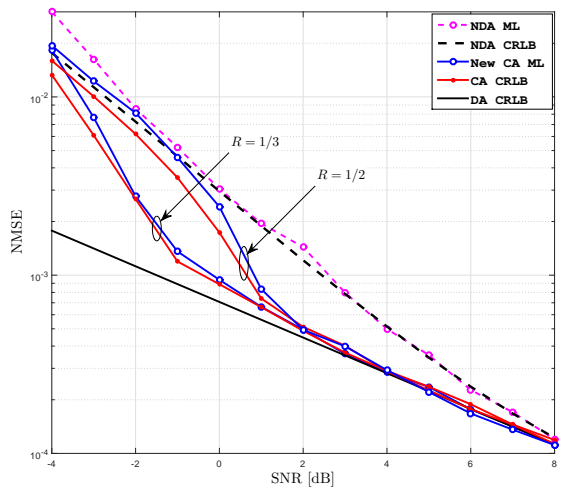}
\end{center} 
\caption{NMSE of the new  CA ML estimator for different code rates, $R$, as function of the SNR: QPSK, rolloff $=0.2$.}
\label{estimator-QPSK} 
\end{figure}

\begin{figure}[!ht]
\begin{center}
\includegraphics[scale = 0.36] {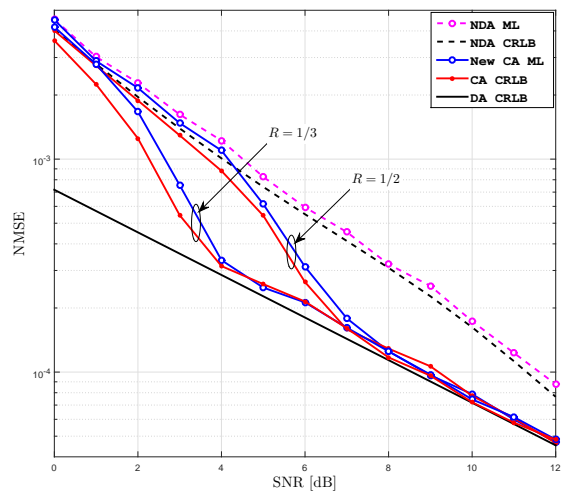}
\end{center}
\caption{NMSE of the new  CA ML estimator for different code rates, $R$, as function of the SNR: 16-QM, rolloff $=0.2$.}
\label{estimator-16-QAM}
\end{figure}

In this section, we provide some graphical representations of the new TD CA CRLBs  for different modulation orders and different coding rates. We also analyze its computational complexity and compare it to that of the existing sum-product expectation-maximization  (SP-EM) timing recovery algorithm [\ref{TSP_Herzet_joint_sum_product_EM}]. The encoder is composed of two identical RSCs concatenated in parallel, having generator polynomials (1,0,1,1) and (1,1,0,1), and a systematic rate $R_0=\frac{1}{2}$ each. The output of the turbo encoder is punctured in order to achieve the desired code rate $R$. For the tailing bits, the size of the RSC encoders memory is fixed to $4$.  We  consider a root-raised-cosine (RRC) signal with roll-off factor $\alpha=0.2$. We  also consider QPSK and 16-QAM, as two representative examples of square-QAM constellations,  and two different coding rates, namely $R=\frac{1}{2}$ and  $R=\frac{1}{3}$.
\begin{figure}[!h]
 \vskip 0.25 cm 
\begin{center}
\includegraphics[scale = 0.35] {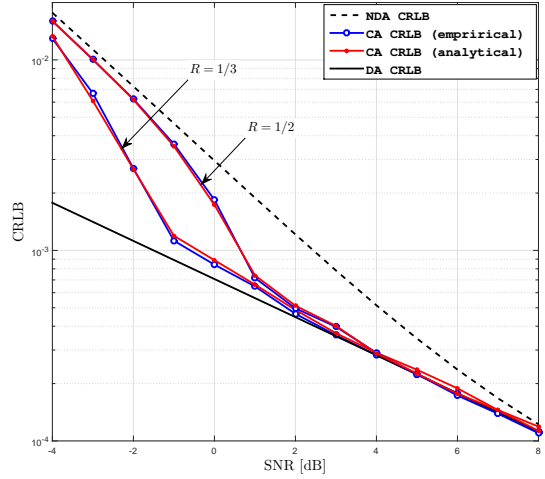}
\end{center}  
\caption{Comparison between the empirical and analytical CA CRLBs for different code rates, $R$, as function of the SNR: QPSK, rolloff $=0.2$.}
\label{crlb-QPSK}
\end{figure}
 \\
We begin by verifying in Figs. \ref{crlb-QPSK} and \ref{crlb-16-QAM} that the new analytical CA CRLBs  coincide with their \textit{empirical} counterparts obtained previously in [\ref{EURASIP_noels_1}]  from exhaustive Monte-Carlo simulations. In fact,  unlike our closed-form  solution, an extremely large number of noisy observations  was generated in [\ref{EURASIP_noels_1}] in order to find an empirical value for the expectation involved in the Fisher information (\ref{Fisher_dinformation_definition}). Hence, our new analytical expression corroborates these previous attempts to evaluate the underlying TD CA CRLBs \textit{empirically} and allow their immediate evaluation for any square-QAM turbo-coded signal.
\\As expected, we also see from both figures that the CA CRLBs are smaller than their NDA counterparts. This highlights the performance improvements that can be achieved by a coded system over an uncoded one by exploiting the information about the transmitted bits that is obtained from the SISO decoders. Additionally and most prominently, the CA CRLBs decrease rapidly and reach the DA CRLBs  which are the best bounds  ever one would be able to achieve if all the transmitted symbols were perfectly known to the receiver, hypothetically.
\\In the sequel, we also assess the performance of the new TD CA ML estimator using the normalized (by $T^2$) mean square error (NMSE) as a performance measure:
\begin{eqnarray}
\textrm{NMSE}&=&\frac{1}{T^2}\frac{\sum_{m=1}^{M_c}\left(\widehat{\tau}_{\textrm{ML-CA}}^{[m]}-\bar{\tau}\right)^2}{M_c},
\end{eqnarray} 
where  $\widehat{\tau}_{\textrm{ML-CA}}^{[m]}$ is the estimate of $\tau$ generated from the $m^{th}$ Monte-Carlo run for $m=1,2\ldots,M_c$. In Figs. \ref{estimator-QPSK} and \ref{estimator-16-QAM}, we plot the NMSE of the new estimator for QPSK and 16-QAM transmissions obtained from $M_c=5000$ Monte-Carlo trials, and  benchmark the resulting performance curves against the corresponding new CA CRLBs. To illustrate the performance advantage  brought   by CA estimation as compared to NCA estimation (from the algorithmic point of view), we also plot in the same figures the NMSE of the NDA TD ML estimator (\ref{E24}).
Figs. \ref{estimator-QPSK} and \ref{estimator-16-QAM} show that the potential estimation performance gains (attributed to the decoder's assistance) made predictable now \textit{theoretically} by the CA CRLBs can be achieved practically by the newly proposed CA ML estimator. More interestingly, the new estimator almost reaches the CA CRLB over the entire practical SNR range confirming thereby its statistical efficiency.  
\\In the same figures, we can also observe unambiguously the effect of the coding rate, $R$, on CA estimation performance. Even though the same NMSE levels are achieved at relatively high SNRs for $R=\frac{1}{2}$ and $R=\frac{1}{3}$, the estimator performs quite differently for the two rates at the same  SNR values. In fact, with  smaller coding rates, more redundancy is introduced by the encoder and, hence, the decoder becomes more likely able to correctly detect the transmitted bits, thereby enhancing  the estimation performance. Now, if we turn the tables and assess the effect of  modulation order on  estimation performance at the same coding rate, we observe without any surprise that it deteriorates with larger constellations at any given  SNR level.  This  typical behavior was already observed in NDA estimation and, as a matter of fact, in any parameter estimation problem involving linearly-modulated signals. Indeed, when the modulation order increases, the inter-symbol distance decreases for normalized-energy constellations. As such, at the same SNR level, noise components  have a relatively worse impact on symbol detection and parameter estimation in general. 
 \begin{figure}[!ht]
\vskip -0.2 cm
\begin{center}
\includegraphics[scale = 0.4] {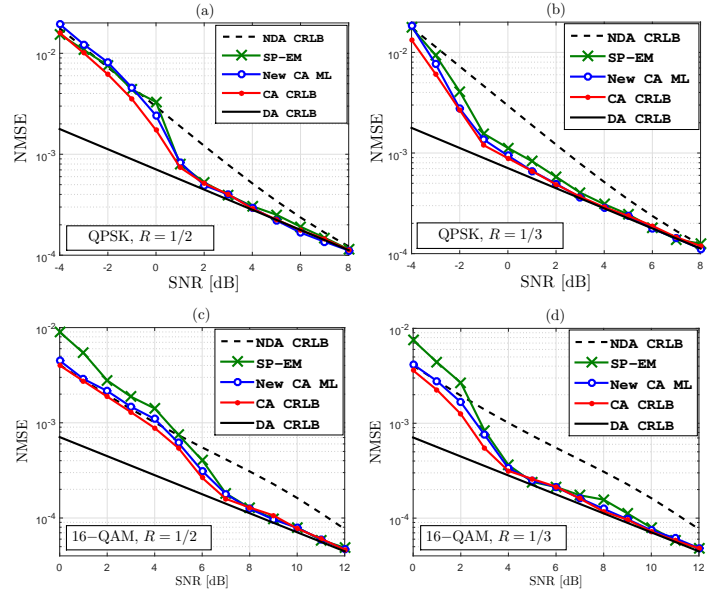}
\end{center}
\caption{NMSE of the new  CA ML estimator and SP-EM for different code rates and modulation orders, rolloff $=0.2$.}
\label{SPEM-subplots}
\end{figure}
\begin{figure}[!ht]
\begin{center}
\includegraphics[scale = 0.41] {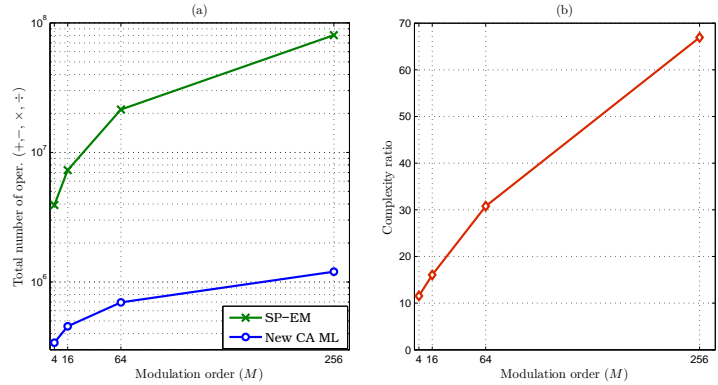}
\end{center}
\caption{complexity  of the new  CA ML estimator and SP-EM for different code rates versus the modulation order: (a) total number of operations, and (b) complexity ratio.}
\label{SPEM-complexity}
\end{figure} 
 \par\indent Finally, we compare the new CA ML TDE to the existing SP-EM ML-based algorithm both in terms of estimation performance and computational complexity in Figs. \ref{SPEM-subplots} and \ref{SPEM-complexity}, respectively.\\
In Fig. \ref{SPEM-subplots}, even though both estimators perform nearly the same with QPSK signals over the entire SNR range, we observe with  16-QAM a clear advantage of the new CA ML TD estimator over SP-EM at low SNR levels.  The superiority of the  proposed estimator over SP-EM can be even better appreciated when it comes to computational complexity. In fact, we plot in Fig. \ref{SPEM-complexity}-(a) the total number of operations (i.e., additions, multiplications, and divisions) required by both estimators versus the modulation order. There we can see that the new CA ML estimator entails much lower computational load. The ratio of complexities depicted in Fig. \ref{SPEM-complexity}-(b) suggests, indeed, that the proposed estimator is about 30 and 70 times computationally less  expensive than SP-EM for 64- and 256-QAM, respectively.      
\section{Conclusion}\label{section_6}
In this paper, we derived for the first time the closed-form expressions of the Cramér-Rao lower bounds for code-aided symbol timing estimation from turbo-coded square-QAM transmissions. The new CA CRLBs revealed the huge performance improvements in terms of timing recovery  are achievable by exploiting the soft information delivered by  SISO decoders at each turbo iteration. The new analytical CRLBs coincide exactly with their empirical counterparts established in previous pioneering works on the subject but from exhaustive Monte-Carlo simulations. We also developed a new code-aided ML time delay estimator that is able to achieve the potential performance gains made thoroughly and instantly predictable by the new  closed-form CA CRLBs. The new estimator also exhibits a remarkable  advantage in terms of computational complexity as compared to the most powerful ML-type algorithm that exists in the literature, namely SP-EM. 
  Simulations results also show, as intuitively expected, that the CA estimation performance  improves by decreasing the coding rate, i.e., increasing the amount of redundancy. 
\section* {Appendix A}
\noindent \underline{A.1) Proof of \textsc{Lemma 1}}:\\~\\
In order to find the pdfs of  $u_k(\tau)$ and $v_i(\tau)$ defined in (\ref{inphase}) and (\ref{quadrature}), respectively, and  prove that they are two independent RVs,  we define the following proper complex RV:
\begin{eqnarray}\label{y_filterd}
\!\!\!\!\!\!\!\!\!\!\!\!y_k(\tau)&\!\!\!\!\triangleq\!\!\!\!&\int_{-\infty}^{+\infty}\!\!\!y(t)~h(t-kT-\tau)dt~=~u_k(\tau)+jv_k(\tau),
\end{eqnarray}
which verifies  $p\big[y_k(\tau)\big] = p\big[u_k(\tau),v_k(\tau)\big]$. Moreover, replacing $y(t)$ by its expression given by (\ref{Output}) in (\ref{y_filterd}) and  resorting to some easy algebraic manipulations,  we obtain:
\begin{eqnarray}\label{z_k_explicit}
y_k(\tau)=\sqrt{E_s}\sum_{k'=1}^{K}a(k')\underbrace{\int_{-\infty}^{+\infty}\!\!\!h(x)h\big(x+[k'-k]T\big)dt}_{g([k'-k]T)}+~w_k(\tau),\nonumber
\end{eqnarray}
where $w_k(\tau)$ is the filtered noise component, i.e.:
\begin{equation}\label{filtered_noise}
w_k(\tau)\triangleq\int_{-\infty}^{+\infty}\!\!\!w(t)h(t-kT-\tau)dt.
\end{equation}
Recall that the shaping pulse $g(t)$ verifies the first Nyquist criterion stated in (\ref{nyquist_condition}), i.e., $g([k'-k]T)=\delta(k'-k)$,  thereby leading to:
\begin{eqnarray}\label{z_k_simplified}
y_k(\tau)=\sqrt{\mathsmaller{E_s}}a(k)+~w_k(\tau).
\end{eqnarray}
Further, it can be  verified from (\ref{filtered_noise}) that $w_k(\tau)$ is Gaussian distributed with zero-mean and  variance $2\sigma^2$. Hence, the pdf of $y_k(\tau)$  conditioned on $a(k)$ is also Gaussian; i.e., $\forall c_m\in\mathcal{C}_p$ we have:
\begin{eqnarray}\label{pdf_z_1}
\!\!\!\!\!\!\!\!p\big[y_k(\tau)|a(k)=c_m\big]&\!\!\!\!=\!\!\!\!&\mathsmaller{\frac{1}{2\pi\sigma^2}}\exp\!\left\{\!-\mathsmaller{\frac{1}{2\sigma^2}}\big|y_k(\tau)\!-\!\mathsmaller{\sqrt{E_s}}c_m\big|^2\!\right\}.\nonumber
\end{eqnarray}
After expanding the modulus in the exponential argument, it can be easily shown that $\forall~c_m\in\mathcal{C}_p$ we have:
\begin{eqnarray}\label{pdf_z_2}
p\big[y_k(\tau)|a(k)=c_m\big]&\!\!\!\!=\!\!\!\!&\mathsmaller{\frac{1}{2\pi\sigma^2}}\mathlarger{e^{-\frac{|y_k(\tau)|^2}{2\sigma^2}}}\Omega_{\tau}\big(c_m, y(t)\big),
\end{eqnarray}
where $\Omega_{\tau}\big(c_m, y(t)\big)$ is given in (\ref{omega_nonaveraged}). Then, by averaging over all the constellation points in $\mathcal{C}_p$ and recalling the expression of $\bar{\Omega}_k(\tau)$ in (\ref{H_definition}), the pdf of $y_k(\tau)$ is obtained as:
\begin{eqnarray}\label{pdf_z_3}
p\big[y_k(\tau)\big]=\mathsmaller{\frac{1}{2\pi\sigma^2}}\mathlarger{e^{-\frac{|y_k(\tau)|^2}{2\sigma^2}}}\bar\Omega_{k}(\tau).
\end{eqnarray}
Finally, using the factorization of $\bar\Omega_{k}(\tau)$ obtained in (\ref{h_tau=F*F}) along with $|y_k(\tau)|^2=u_k^2(\tau)+v_k^2(\tau)$ and  $\beta_k=\beta_{k,2p}\beta_{k,2p-1}$, it follows that:
\begin{eqnarray}\label{pdf_z_4}
\!p\big[y_k(\tau)\big]&=&\mathsmaller{\frac{4\beta_{k,2p}\beta_{k,2p-1}}{2\pi\sigma^2}}\mathlarger{e^{-\frac{u_k^2(\tau)+v_k^2(\tau)}{2\sigma^2}}}F_{k,2p}\big(u_k(\tau)\big)F_{k,2p-1}\big(v_k(\tau)\big.)\nonumber\\
\!&=&\underbrace{\mathsmaller{\frac{2\beta_{k,2p}}{\sqrt{2\pi\sigma^2}}}\mathlarger{e^{-\frac{u_k^2(\tau)}{2\sigma^2}}}\!F_{k,2p}\big(u_k(\tau)\big)}_{p[u_k(\tau)]}~\!\!\underbrace{\mathsmaller{\frac{2\beta_{k,2p-1}}{\sqrt{2\pi\sigma^2}}}\mathlarger{e^{-\frac{v_k^2(\tau)}{2\sigma^2}}}\!F_{k,2p-1}\!\big(v_k(\tau)\big)}_{p[v_k(\tau)]}\!.\nonumber
\end{eqnarray}
From the last equality, we obtain $p\big[y_k(\tau)\big]=p\big[u_k(\tau)\big]p\big[v_k(\tau)\big]$. But since from (\ref{y_filterd}) we already have $y_k(\tau)=u_k(\tau)+jv_k(\tau)$, then we also have $p\big[y_k(\tau)\big]=p\big[u_k(\tau),v_k(\tau)\big]$. Therefore, it follows that $p\big[u_k(\tau),v_k(\tau)\big]=p\big[u_k(\tau)\big]p\big[v_k(\tau)\big]$, meaning that the two RVs $u_k(\tau)$ and $v_k(\tau)$ are actually independent and their distributions are, respectively, given by (\ref{pdf_u}) and (\ref{pdf_v}).
\\~\\
\noindent \underline{A.2) Statistical Independence of $u_k(\tau)$ and $\bm{\dot}u_k(\tau)$}:\\ ~\\
First, it follows from (\ref{y_filterd}) that:
\begin{eqnarray}
\!\!\!\!\!\!\!\!\!\!\!\!\bm{\dot}u_k(\tau)&\!\!\!\!=\!\!\!\!&\frac{\partial\Re\{y_k(\tau)\}}{\partial\tau}=-\int_{\mathbb{R}}\!\Re\big\{y(t)\big\}~\bm\dot h(t-kT-\tau)dt.
\end{eqnarray}
Again, we replace $y(t)$ by its expression given in (\ref{Output}) and then we use the fact that $\bm\dot{g}(.)$ is an odd function to show that:
\begin{eqnarray}\label{u_dot_Appendix_A}
\!\!\!\!\!\!\!\!\!\!\!\!\bm{\dot}u_k(\tau)&\!\!\!\!=\!\!\!\!&\sqrt{E_s} \sum\displaylimits_{k'=1}^K \Re\big\{a(k')\big\}\bm \dot g\big([k-k']T\big)+\Re\big\{\bm{\dot}{w}_k(\tau)\big\},
\end{eqnarray}
where $\bm\dot w_k(\tau)$ is the derivative of $w_k(\tau)$   with respect to $\tau$, which is obtained by replacing $h(t-kT-\tau)$ by $-\bm\dot{h}(t-kT-\tau)$ back in  (\ref{filtered_noise}). Recall also that $\bm \dot g(0)=0$ \big(since the maximum of $g(x)$ is located at $0$), leading to:
\begin{eqnarray}\label{dot_u_k}
\!\!\!\!\!\!\!\!\!\!\!\!\bm{\dot}u_k(\tau)&\!\!\!\!=\!\!\!\!& \sqrt{E_s} \sum\displaylimits_{\mathclap{\substack{k'=1\\k'\neq k}} }^K \Re\big\{a(k')\big\}\bm \dot g\big([k-k']T\big)+\Re\big\{\bm{\dot}{w}_k(\tau)\big\}.
\end{eqnarray}
Recall also from (\ref{y_filterd}) that $u_k(\tau)=\Re\{y_k(\tau)\}$ and, therefore, we have from (\ref{z_k_simplified}) :
\begin{equation}\label{u_k}
u_k(\tau)= \sqrt{E_s} \Re\big\{a(k)\big\}+\Re\big\{w_k(\tau)\big\}.
\end{equation}
Notice from  (\ref{dot_u_k})  that $\bm{\dot}u_k(\tau)$ involves the contribution of all the symbols except the $k^{th}$ one \big[i.e., $a(k)$\big] that is, in turn, the only one involved in $u_k(\tau)$ as seen from (\ref{u_k}). Since the symbols are mutually independent, then in order to show the independence of $u_k(\tau)$ and  $\bm{\dot}u_k(\tau)$, it suffices to show the independence of $w_k(\tau)$ and $\bm{\dot}{w}_k(\tau)$. These are actually two RVs that are obtained from  linear transformations (i.e., integral and derivative) of the same Gaussian process $w(t)$ and, hence, they are also Gaussian distributed. Their   cross-correlation is given by: 
\begin{eqnarray}\label{E_U_dU}
\!\!\mathbb{E}\left\{w_k(\tau) \bm \dot w_k(\tau)\right\}\!&=&\iint_{-\infty}^{+\infty}\!\!\!\!\mathbb{E}\big\{w(t_1)w(t_2)\big\}
h(t_1-kT-\tau)\bm{\dot}h(t_2-kT-\tau)dt_1dt_2\nonumber\\
\!\!&=&2\sigma^2\!\iint_{-\infty}^{+\infty}\!\!\!\!\delta(t_1-t_2)h(t_1)\bm{\dot}h(t_2)dt_1dt_2\nonumber\\
\!\!&=&2\sigma^2\bm \dot g(0)\nonumber\\
\!\!&=&0,
\end{eqnarray}
meaning that the two Gaussian-distributed RVs $w_k(\tau)$ and $\bm{\dot}{w}_k(\tau)$ are uncorrelated and, therefore,  independent as well. Consequently, $u_k(\tau)$ and $\bm{\dot}u_k(\tau)$ are also independent. 
\section*{Appendix B} 
Using the decomposition $\mathcal{C}_p=\widetilde{\mathcal{C}}_p\cup(-\widetilde{\mathcal{C}}_p)\cup\widetilde{\mathcal{C}}_p^*\cup(-\widetilde{\mathcal{C}}_p^*)$
 and noticing that:
 \begin{equation}
 \Re\big\{\tilde{c}_m\big\}^2=\Re\big\{\!-\tilde{c}_m\!\big\}^2=\Re\big\{\tilde{c}^*_m\big\}^2=\Re\big\{\!-\tilde{c}^*_m\big\}^2,~~~~\forall~\tilde{c}_m\in \widetilde{\mathcal{C}}_p,\nonumber
 \end{equation}
%
  we rewrite (\ref{whole-alphabet}) as follows:
\begin{eqnarray}\label{single_sum_in_top_right}
\!\!\!\!\!\!\!\!\!\!\mathbb{E}\Big\{\!\Re\big\{ a(k)\big\}^2\!\Big\}
&\!\!=\!\!& \sum_{\tilde{c}_m\in\tilde{\mathcal{C}}_p}\!\!\!\Re\big\{\!\tilde{c}_m\!\big\}^2\Big(\!P\big[a(k)\!=\!\tilde{c}_m\big]\!+\!P\big[a(k)\!=\!-\tilde{c}_m\big]+P\big[a(k)\!=\!\tilde{c}^*_m\big]\!+\!P\big[a(k)\!=\!-\tilde{c}^*_m\big]\!\Big).
\end{eqnarray}
Moreover, by using the explicit expressions of the symbols' APPs given in (\ref{prob_LSB_explicit_1})-(\ref{prob_LSB_explicit_4}), along with the identity   $\cosh(x)+\cosh(y)=2\cosh(\frac{x+y}{2})\cosh(\frac{x-y}{2})$, we obtain:
\begin{eqnarray}
Pr\big[a(k)\!=\!\tilde{c}_m\big]\!+\!Pr\big[a(k)\!=\!-\tilde{c}_m\big]\!+\! Pr\big[a(k)\!=\!\tilde{c}^*_m\big]\!+\!Pr\big[a(k)\!=\!-\tilde{c}^*_m\big]\!\!&\!\!\!\! &\!\!\nonumber\\
&\!\!\!\!&\!\!\!\!\!\!\!\!\!\!\!\!\!\!\!\!\!\!\!\!\!\!\!\!\!\!\!\!\!\!\!\!\!\!\!\!\!\!\!\!\!\!\!\!\!\!\!\!\!\!\!\!\!\!\!\!\!\!\!\!\!\!\!\!\!\!\!\!\!\!\!\!= ~ 2\beta_k\mu_{k,p}(\tilde{c}_m)\!\left[\mathsmaller{\cosh\left({\frac{L_{2p}(k)+L_{2p-1}(k)}{2}}\right)}\!+\!\mathsmaller{\cosh\left({\frac{L_{2p}(k)-L_{2p-1}(k)}{2}}\right) }\right],\nonumber\\
&\!\!\!\!&\!\!\!\!\!\!\!\!\!\!\!\!\!\!\!\!\!\!\!\!\!\!\!\!\!\!\!\!\!\!\!\!\!\!\!\!\!\!\!\!\!\!\!\!\!\!\!\!\!\!\!\!\!\!\!\!\!\!\!\!\!\!\!\!\!\!\!\!\!\!\!\!=~ 4\beta_k\mu_{k,p}(\tilde{c}_m) \mathsmaller{\cosh\left({\frac{L_{2p}(k) }{2}}\right)}\mathsmaller{\cosh\left({\frac{ L_{2p-1}(k)}{2}}\right)},\label{app_0}
\end{eqnarray} 
%
 Now, plugging (\ref{app_0}) back into (\ref{single_sum_in_top_right}), rewriting the sum over $\tilde{c}_m\in\tilde{\mathcal{C}}_p$ as a double sum over the counters $i$ and $n$ [where $\tilde{c}_m=(2i-1)d_p+j(2n-1)d_p$ as done in (\ref{H_expression_4})], and using the decomposition in  (\ref{mu_c_m_factorized}), it can be shown that:
\begin{eqnarray}\label{result}
\!\!\!\!\!\!\mathbb{E}\Big\{\Re\big\{a(k)\!\big\}^2\Big\}&=&4\beta_k \sum_{i=1}^{2^{p-1}}\sum_{n=1}^{2^{p-1}}\Big[(2i-1)^2d_p^2\theta_{k,2p}(i)\theta_{k,2p-1}(n)\times\mathsmaller{\cosh\left({\frac{L_{2p}(k) }{2}}\right)\cosh\left({\frac{ L_{2p-1}(k)}{2}}\right)}\Big]\nonumber\\
&=&2\beta_{k,2p}\mathsmaller{\cosh\left(\!{\frac{ L_{2p}(k)}{2}}\!\right)}\sum_{i=1}^{2^{p-1}}(2i-1)^2d_p^2\theta_{k,2p}(i) \times2\beta_{k,2p-1}\mathsmaller{\cosh\left(\!{\frac{L_{2p-1}(k) }{2}}\!\right)}\sum_{n=1}^{2^{p-1}}\theta_{k,2p-1}(n),\nonumber\\&&
\end{eqnarray}
where the decomposition $\beta_k=\beta_{k,2p}\beta_{k,2p-1}$ was used in the last equality as well.
Moreover, it has been recently shown in  [\ref{Bellili-CA-CFO}, \textsc{lemma 3}] that for $q=2p$ and $2p-1$:
\begin{eqnarray}\label{simplification}
2\beta_{k,q} \mathsmaller{\cosh\Big(\frac{L_{q}(k)}{2}\Big)}\displaystyle \sum_{n=1}^{2^{p-1}}\theta_{k,q} {(n)}&=&1,
\end{eqnarray}
which is used back in (\ref{result}) to obtain the following result:
\begin{eqnarray}\label{first_expectation}
\mathbb{E}\bigg\{\Re\big\{a(k)\!\big\}^2\bigg\}&=& \omega_{k,2p}.
\end{eqnarray}

\end{document}